\documentclass[12pt,a4paper]{article}
\usepackage{amssymb} 
\usepackage{amsmath}
\usepackage{mathtools}
\usepackage{amsfonts}    
\usepackage{dsfont}
\usepackage{pdfpages}
\usepackage{verbatim}
\usepackage{tensor}
\usepackage{hyperref}
\usepackage{mathrsfs}
\usepackage[mathscr]{euscript}
\usepackage{tikz}
\usepackage[utf8]{inputenc}
\usepackage[symbol]{footmisc}

\usepackage{tikz-cd}
\usepackage{extarrows}

\newcommand{\ba}{\begin{align}}

\newcommand{\be}{\begin{equation}}
\newcommand{\ee}{\end{equation}}
\def\bd{\begin{tikzpicture}}
\def\ed{\end{tikzpicture}}

\DeclareMathOperator{\SL}{SL}

\DeclareMathOperator{\SU}{SU}

\DeclareMathOperator{\Ima}{Im}

\DeclareMathOperator*{\OplusInt}{%
\mathchoice%
  {\ooalign{$\displaystyle\oplus$\cr\hidewidth$\displaystyle\int$\hidewidth\cr}}
  {\ooalign{\raisebox{.14\height}{\scalebox{.7}{$\textstyle\oplus$}}\cr\hidewidth$\textstyle\int$\hidewidth\cr}}
  {\ooalign{\raisebox{.2\height}{\scalebox{.6}{$\scriptstyle\oplus$}}\cr$\scriptstyle\int$\cr}}
  {\ooalign{\raisebox{.2\height}{\scalebox{.6}{$\scriptstyle\oplus$}}\cr$\scriptstyle\int$\cr}}
}

\newcommand{\interior}[1]{%
  {\kern0pt#1}^{\mathrm{o}}%
}

\advance\voffset by -1.5cm
\advance\hoffset by -0.8cm
\renewcommand{\thefootnote}{\arabic{footnote}}

\newcommand{\symfootnote}[1]{%
\let\oldthefootnote=\thefootnote%
\stepcounter{mpfootnote}%
\addtocounter{footnote}{-1}%
\renewcommand{\thefootnote}{\fnsymbol{mpfootnote}}%
\footnote{#1}%
\let\thefootnote=\oldthefootnote%
}

\allowdisplaybreaks[1]

\begin{document}
\vspace*{-5\baselineskip}%
\hspace*{\fill} \mbox{\footnotesize{\textsc{MPP-2024-196}}}

\vspace*{1.5cm}

\begin{center}
{\Large 
{\bf The $\mathfrak{su}(2)_{-1}$ WZW model}}
\vspace{2.5cm}

\setcounter{mpfootnote}{0}

\vspace{0.5cm}
{\large Elia Mazzucchelli}\symfootnote{{\tt E-mail: eliam@mpp.mpg.de}} \\
Max-Plank-Institut f\"ur Physik, Werner-Heisenberg-Institut,\\
D-85748 M\"unchen, Germany \\
\vspace*{0.5cm}

{\bf Abstract}
\end{center}

Some WZW models on affine Lie superalgebras at critical level describe string theory on AdS backgrounds at critical values of the string tension. This is the case of $\mathfrak{psu}(1,1|2)_1$ for ${\rm AdS}_3 \times {\rm S}^3$ and potentially of $\mathfrak{u}(2|2)_1$ (or related algebras) for ${\rm AdS}_5 \times {\rm S}^5$. Many interesting features of these superalgebra models are already captured by their affine subalgebra $\mathfrak{su}(2)_{-1}$. In this paper we study the WZW model on $\mathfrak{su}(2)_{-1}$: we classify the representations, introduce a free field realisation, and decompose the free field modules in terms of $\mathfrak{su}(2)_{-1}$. We find continuous and discrete modular invariants and see that the latter naturally leads to considering superalgebra extensions of $\mathfrak{su}(2)_{-1}$. Lastly, we find an invariant for the free field theory of four symplectic bosons.

\newpage
\renewcommand{\theequation}{\arabic{section}.\arabic{equation}}

\section{Introduction} \label{sec:intro}

The tensionless limit of string theory, particularly in the context of the AdS/CFT correspondence, has garnered significant attention in recent years. A promising approach to understanding the dynamics of tensionless strings, especially in AdS spacetimes, is through Wess-Zumino-Witten (WZW) models \cite{Maldacena:2000hw}. 

One important result in this direction comes from a recent exact solvable AdS/CFT duality where strings on ${\rm AdS}_3 \times {\rm S}^{3} \times \mathbb{T}^4$ with $k=1$ unit of NS-NS flux are dual to the symmetric orbifold of $\mathbb{T}^{4}$ \cite{Eberhardt:2018ouy,Eberhardt:2019ywk}. In this case the worldsheet theory on ${\rm AdS}_3 \times {\rm S}^{3}$ is governed by a $\mathfrak{psu}(1,1|2)_{k}$ WZW model at level $k=1$. The representations of this model have been instrumental in elucidating how stringy excitations emerge in the low-energy limit: the absence of the long string continuum is essentially because a shortening condition (null vector) at $k = 1$ removes the continuous representations of $\mathfrak{sl}(2,\mathbb{R})_1$ except
for the bottom of the continuum.

Later, a similar construction for ${\rm AdS}_5 \times {\rm S}^5$ has been proposed, at the point in moduli space where the string theory is dual to free super Yang Mills (SYM) \cite{Gaberdiel:2021qbb,Gaberdiel:2021jrv}. The worldsheet theory involves a WZW model based on $\mathfrak{psu}(2,2|4)_k$, again at level $k=1$. According to this proposal, the physical state condition must remove half of the symmetry (one copy of the symmetry algebra $\mathfrak{psu}(2,2|4) \oplus \mathfrak{psu}(2,2|4)$ coming from the WZW model description). In fact, the ${\cal N}=4$ superconformal symmetry in 4D is described by a single copy of $\mathfrak{psu}(2,2|4)$. Motivated by this seemingly excessive gauging and by the integrability approach \cite{Beisert:2005tm}, where the natural description is in terms of $\mathfrak{su}(2|2)$, in a companion paper we explored the WZW model on $\mathfrak{u}(2|2)_1$ \cite{Gaberdiel:2023nhb}. The latter furnishes a potential candidate for a more `economical' description of tensionless strings on ${\rm AdS}_5 \times {\rm S}^5$. Even though we did not manage to match the ${\cal N}=4$ free SYM spectrum to any modular invariant spectrum of $\mathfrak{u}(2|2)_1$ yet, the former is naturally embedded in both continuous and discrete invariants of the latter. We therefore can not exclude that the physical spectra can be matched by the right implementation of physical state conditions and including appropriate ghost contributions to the WZW partition function.

\smallskip

In both worldsheet descriptions of ${\rm AdS}_{3} \times {\rm S}^3$ and ${\rm AdS}_{5} \times {\rm S}^5$ mentioned above, the corresponding WZW model is on a superalgebra that contains a non-compact factor\footnote{Note that $\mathfrak{psu}(1,1|2)_1$ actually contains a factor $\mathfrak{sl}(2,\mathbb{R})_1$ which is not isomorphic to $\mathfrak{su}(2)_{-1}$ as a real affine Lie algebra. Nevertheless, they both possess the same complexification and hence, from a representation theory point of view, we can ignore the distinction. The reason for favoring the notation $\mathfrak{su}(2)_{-1}$ in this paper is that we will be mainly interested in discrete WZW spectra. The latter can be interpreted as quantising the magnetic quantum number, and the resulting theory can be understood as coming from a sigma model on the (compact) Lie group ${\rm SU}(2)$.} $\mathfrak{su}(2)_{-1}$. The $\mathfrak{su}(2)_{-1}$ WZW captures in a sense many of the interesting features of the WZW on the respective superalgebras. This is because for both $\mathfrak{psu}(2|2)_1$ and $\mathfrak{u}(2|2)_1$, the inclusion of the respective bosonic subalgebras yields a conformal embedding \cite{Conf_emb}. 
For $k \in \mathbb{N}$ the $\mathfrak{su}(2)_{k}$ model is integrable and constitutes a rational conformal field theory (CFT). However, for $k=-1$ the model falls out of this class. The set of admissible representations is much less constrained and richer: it is continuously parameterized by the (real) value of the Casimir, and the modules come in three families (those associated with finite-dimensional, discrete, or continuous $\mathfrak{su}(2)$ representations). Another important distinctive feature from the integrable case is that the spectral flow automorphism has not finite order and gives rise to additional (infinitely many) modules. All of these features directly transfer to the superalgebra WZW models mentioned previously.

\smallskip

In addition to the string theory applications, the WZW model on $\mathfrak{su}(2)_{-1}$ has interesting mathematical properties of its own. In fact, it yields one of the simplest examples of non-unitary CFT, having negative central charge $c=-3$. Theories with negative central charge arise for instance in ghost systems \cite{Ridout:2008nh}. The $\mathfrak{su}(2)_{-1}$ model also provides an instance of a logarithmic CFT, see \cite{Flohr:2001zs,Gaberdiel:2001tr,Creutzig:2013hma} for a review. Non-unitary logarithmic CFTs play an important role in Fishnet theories \cite{Gurdogan:2015csr,Kazakov:2022dbd}. In particular, the spectrum contains reducible but indecomposable representations, leading to logarithmic correlation functions, see \cite{Schomerus:2005bf,Gotz:2006qp,Saleur:2006tf}. The characters present convergence issues observed in other fractional level models \cite{Flohr:1995ea,Creutzig:2012sd,Creutzig:2013yca} and the fusion rules are expected to have a complicated structure \cite{Ridout:2010jk,Gaberdiel:2001ny}.

\smallskip

The focus of this paper is on the representation theory and modular invariants of the $\mathfrak{su}(2)_{-1}$ WZW model, as well as its free field realisation. We summarize our contributions. In Section \ref{section_affine_representations} we use the Kac-Kazhdan determinant on $\mathfrak{su}(2)_{-1}$ Verma modules to classify all singular vectors. We find that the singular submodule of a Verma module associated to a highest/lowest weight discrete representation with $j \in \pm \tfrac{1}{2}\mathbb{N}$, or to a continuous representations with $j-\lambda \in \mathbb{Z}$, is generated by a single null vector (\ref{sing_vector_D+}), (\ref{sing_vector_D-}). See Section (\ref{su2_representations}) for our conventions on $\mathfrak{su}(2)$ representations. The $\mathfrak{su}(2)_{-1}$ model is known to have a free field realisation in terms of four symplectic bosons~\cite{Goddard:1987td}. We compute the characters of both the $\mathfrak{su}(2)_{-1}$ and free field modules, and decompose the latter in terms of the former. We find an interesting decomposition of the free field representations in terms of $\mathfrak{su}(2)_{-1}$ irreducibles, such as (\ref{R_decomp}) and~(\ref{NS_decomposition}). 

We compute the modular transformations of $\mathfrak{su}(2)_{-1}$ characters in Section \ref{section_fin_dim_discr} and~\ref{sec_cont_repr}. This allows us to determine a set of modular invariant partition functions; we find invariant spectra involving only discrete representations (\ref{discr_spectra}) and only continuous representations (\ref{cont_diag_spectrum}), which are analogous to those found in \cite{Maldacena:2000hw} in the study of $\mathfrak{sl}(2,\mathbb{R})$ WZW models. More interestingly, we find two modular invariant partition functions (\ref{part_functs}) involving only a discrete subset of $\mathfrak{su}(2)_{-1}$ characters. In Section (\ref{sect_inv_from_coset}) we give two possible interpretation for these two partition functions: as deriving from cosets constructions of $\mathfrak{psu}(2|2)_1$ spectra (\ref{psu22_spectra}) found \cite{Gaberdiel:2023nhb}, or as simple current (super)algebra extensions. According to this last interpretation, the first invariant in (\ref{part_functs}) is associated to the diagonal spectrum (\ref{su2_1_spectrum}) of a superalgebra extension of $\mathfrak{su}(2)_{-1}$ by simple currents (\ref{extended_modules}). We stress that we do not know if such an extended algebra exists as a vertex operator superalgebra, but if it does, it certainly is an unusual algebra due to the presence of states with negative conformal dimension. Interestingly, this extended algebra possesses two irreducible representations (\ref{X_module}), on which the spectral flow action (\ref{sigma}) of $\mathfrak{su}(2)_{-1}$ becomes an involution, and the modular $S$-matrix is the same of that of $\mathfrak{su}(2)_1$ (\ref{su2_1_S_matr}). The second invariant in (\ref{part_functs}) corresponds to the diagonal invariant (\ref{su2_1_spectrum2}) of a bosonic simple current extension of $\mathfrak{su}(2)_{-1}$ by (\ref{extension2}). The extended theory, if it exists, involves eight irreducible representations (\ref{C0_modules}), (\ref{C1_modules}) and their spectrally flowed images, and the $S$-matrix is given by (\ref{S_matrix_interesting}). 

Finally, in Section \ref{section_free_field_invariant} we present a modular invariant partition function (\ref{sp4_inv}) for the free field theory of four symplectic bosons. We stress the difficulty of incorporating the spectral flow action (\ref{sigma}), which has an infinite orbit on the free field modules, at the level of \textit{character functions}. We overcome this difficulty by taking inspiration from the recently proposed free field invariant for four fermions and four symplectic bosons in \cite[(5.21)]{Gaberdiel:2023nhb}, which leads us to consider the spectrum (\ref{proposal}).
We then show that the actual modular invariant partition function is obtained from (\ref{proposal}) by summing over the two contributions associated with the two possible boundary conditions, or spin structures, of the bosonic fields on the torus. This is analogous the free field modular invariant for free fermions on the torus. In \cite{Gaberdiel_2018} the authors point out that it would be interesting to study the decomposition of the free field modules of four symplectic bosons in terms of $\mathfrak{su}(2)_{-1}$ modules, as well as finding a modular invariant for the free field theory. In this paper we therefore answer both questions.

\bigskip

The paper is organized as follows: In Section~\ref{section_su2__1} we define the affine Lie algebra $\mathfrak{su}(2)_k$, recall the form of $\mathfrak{su}(2)$ representations and analyze the singular vectors at $k=-1$.  In Section~\ref{section_free_field_real} we explain a free field realisation for the algebra $\mathfrak{u}(2)_{-1}$, and define the free field representations. Section~\ref{section_characters} explains how to compute the characters of the admissible $\mathfrak{su}(2)_{-1}$ modules and those of the free field representations. We also decompose the free field representations in terms of $\mathfrak{su}(2)_{-1}$ modules. In Section~\ref{section_mod_inv} we compute the modular matrices and identify a number of continuous and discrete modular invariants. Section~\ref{sect_inv_from_coset} is devoted to interpreting the discrete invariants in terms of cosets of $\mathfrak{psu}(2|2)_1$ as well as vertex operator (super)algebra extensions of $\mathfrak{su}(2)_{-1}$. In Section~\ref{section_free_field_invariant} we give a modular invariant for the free field theory of four symplectic bosons. Lastly, Section~\ref{sec:concl} contains the conclusions, and there are two appendices containing details about the free field representations and our conventions on modular functions.

\section{The algebra $\mathfrak{su}(2)_{-1}$ and its representations}
\label{section_su2__1}

The affine Lie algebra $\mathfrak{su}(2)_{k}$ is generated by the modes $J^{a}_{n}$ with $a = \pm, 3$ and $n \in \mathbb{Z}$ satisfying the commutation relations
\begin{equation}\label{su2_k_commutators}
    \begin{aligned}
    [J^{3}_{m}, J^{3}_{n}] &= \tfrac{k}{2}   m  \delta_{m+n,0} \ ,\\
    [J^{3}_{m},J^{\pm}_{n}] &= \pm   J^{\pm}_{m+n} \ , \\
    [J^{+}_{m}, J^{-}_{n}] &= 2   J^{3}_{m+n} + k  m  \delta_{m+n,0} \ .
\end{aligned}
\end{equation}
For $k\neq -2$, the Sugawara construction gives an embedding of the Virasoro algebra in the universal enveloping algebra of $\mathfrak{su}(2)_k$, such that the Virasoro generators $L_n$ are the modes of the Sugawara energy-momentum tensor
\begin{equation}\label{Sugawara_T}
    T^{\mathfrak{su}(2)_k} = \tfrac{1}{k + 2}  \left( J^{3}  J^{3} + \tfrac{1}{2}( J^{+}  J^{-} + J^{-}  J^{+} ) \right) \ ,
\end{equation}
where normal ordering is implicitly assumed. It follows that an affine highest weight state with spin $j$ (with respect to the $\mathfrak{su}(2)$ subalgebra of zero modes) has conformal dimension ($L_0$-eigenvalue) equal to
\begin{equation}\label{h_j}
  h_{j}  = \frac{j(j+1)}{k+2}  \ . 
\end{equation}
Therefore, at fixed level $k$, an affine $\mathfrak{su}(2)_{k}$ weight $(j,k,h_{j})$ is completely determined by the spin $j$ (and similarly for a lowest weight by the replacement $j \mapsto -j$). The central charge of the Virasoro algebra corresponding to (\ref{Sugawara_T}) is
\begin{equation}\label{central_charge}
    c = \frac{3 k}{k+2} \ ,
\end{equation}
which for our case of interest, $k=-1$, takes the value $c=-3$.

The automorphisms of $\mathfrak{su}(2)_{k}$ that preserve the Cartan subalgebra are generated by the \textit{conjugation} automorphism $*$ and the \textit{spectral flow} automorphism $\sigma$. These automorphisms leave the level $k$ invariant and their action is given by 
\begin{align}
\label{conjugation}
        (J^{\pm}_{n})^{*} &= J^{\mp}_{n} \ , \quad \quad \ \ (J^{3}_{n} )^{*} = -J^{3}_{n} \ , \quad \quad \quad \quad \quad \, (L_{n})^{*} = L_{n} \ ,\\\label{sigma}
        \sigma^{w}(J^{\pm}_{n}) &= J^{\pm}_{n \pm w} \ , \quad \sigma^{w}(J^{3}_{n}) = J^{3}_{n} + k  \tfrac{w}{2} \, \delta_{n,0} \ , \quad \sigma^w(L_{n}) = L_{n} +  w  J^{3}_{n} + k  \tfrac{w^2}{4}   \delta_{n,0} \ .
\end{align}
Note that $* \circ \sigma = \sigma^{-1} \circ *$. We define the action of an automorphism $\rho$ of $\mathfrak{su}(2)_{k}$ on an $\mathfrak{su}(2)_k$ module $\mathcal{H}$ as the module $\rho (\mathcal{H} )$ spanned by the states $\bigl[\Phi\bigl]^{\rho}$,
where $\Phi \in \mathcal{H}$ (thus, as vector spaces the two are isomorphic), and the twisted action of the $\mathfrak{su}(2)_k$ generators $A_n$ is defined by
\begin{equation}\label{spec_flo_action}
    A_{n} \, \bigl[\Phi \bigl]^{\rho} := \bigl[ \rho (A_{n} ) \, \Phi\bigl]^{\rho} \ .
\end{equation}
We call the modules $\mathcal{H}^{*}$ and $\sigma^{w}\bigl(\mathcal{H}\bigl)$, for $w \in \mathbb{Z}$, the conjugate and $w$’th spectrally flowed modules of $\mathcal{H}$, respectively.

\subsection{The representations of $\mathfrak{su}(2)$}
\label{su2_representations}

Let us recall some elements of the representation theory of $\mathfrak{su}(2)$. Since $\mathfrak{su}(2)$ is compact, every unitary irreducible representation is finite-dimensional. In the affine case, for $k$ positive integer, $k \in \mathbb{N}$, the $\mathfrak{su}(2)_{k}$ model is integrable and it possesses only $k+1$ unitary integrable highest weight representations, which are characterized by the finite-dimensional spin $\ell \in \tfrac{1}{2}\mathbb{N}_0$ representations of $\mathfrak{su}(2)$ ($\ell \leq \tfrac{k}{2}$) with respect to which the highest weight states transform. However, the $\mathfrak{su}(2)_{-k}$ model is non-integrable, and thus no such restriction applies for it. In particular, it possesses no unitary highest weight representations and the spectrum is continuous.

Even though the spectrum of the $\mathfrak{su}(2)_{-k}$ theory is continuous, from the perspective of the Lie group, we expect the compactness of $\SU(2)$ to constrain the set of allowed representations to a discrete subset characterized by the property that the magnetic quantum numbers are quantised. Nevertheless, the finite $\mathfrak{su}(2)$ representations lying at the highest weight $\mathfrak{su}(2)_{-k}$ representations are allowed to be non-unitary, i.e. infinite-dimensional. We thus look at all possible representations of $\mathfrak{su}(2)$, or equivalently (if disregarding unitarity), of $\mathfrak{sl}(2, \mathbb{R})$. These are classified in the following three families.
\begin{itemize}
    \item The \textit{finite-dimensional} representations $H_{j}$ of spin $j \in \tfrac{1}{2}\mathbb{N}_0$. These are the usual unitary representations of dimension $2 j+1$, characterized by the spin $j$. The Casimir of these representations is 
    \begin{equation}
        C^{\mathfrak{su}(2)}(H_{j} ) = j (j+1) \ .
    \end{equation}
    
    \item The \textit{highest/lowest weight discrete} representations $D^{\pm}_{j}$ of spin $j \in \mathbb{R}$. These are infinite-dimensional non-unitary representations defined by a highest/lowest weight state $|j\rangle$ such that 
    \begin{equation}
        D^{\pm}_{j} ~:~~~~J^{\pm} \, |j\rangle = 0~~~~\text{and}~~~~ J^{3} \, |j\rangle = j \, |j\rangle \ ,
    \end{equation}
    and with Casimir equal to
    \begin{equation}
        C^{\mathfrak{su}(2)}(D^{\pm}_{j} ) = j  (j \pm 1) \ .
    \end{equation}
    They are irreducible for $j \notin \mathbb{R} \setminus \pm \tfrac{1}{2}\mathbb{N}_0$ and reducible but indecomposable otherwise: there is a non-split short exact sequence
    \begin{equation}\label{discrete_ses}
    0 \longrightarrow D^{\pm }_{\mp(j+1)} \longrightarrow D^{\pm}_{\pm j} \longrightarrow H_{j} \longrightarrow 0 ~~~~ \forall \, j \in \pm \tfrac{1}{2} \mathbb{N}_0 \ .
\end{equation}

    \item The \textit{continuous} representations $C_{j}^{\lambda}$, for $j \in \mathbb{R}$ and $\lambda \in \mathbb{R}/\mathbb{Z}$. These are infinite-dimensional non-unitary representations that neither contain a highest nor a lowest weight state, and they are characterized by their Casimir 
    \begin{equation}
        C^{\mathfrak{su}(2)}(C_{j}^{\lambda} ) = j  (j+1) \in \mathbb{R} \ ,
    \end{equation}
    as well as the fractional part of the $J^{3}$-eigenvalues $\lambda \in \mathbb{R}/\mathbb{Z} \cong [0,1)$. More specifically, the representation $C_{j}^{\lambda}$ is defined by states $|m \rangle$ with $m \in \mathbb{Z} + \lambda$ such that
    \begin{equation}
        \begin{aligned}\label{su2_continuous_repr}
            J^{3} \, |m\rangle &= m \, |m \rangle \ ,\\ 
            J^{+} \, |m\rangle &=  \bigl( j (j+1) - m  (m+1) \bigl) |m+1 \rangle \ ,\\
            J^{-} \, |m\rangle &= |m-1 \rangle \ .\\           
        \end{aligned}
    \end{equation}
    Notice that $C^{\lambda}_{j} = C^{\lambda}_{-j-1}$, and hence we may assume that $j \geq -1/2$. Moreover, in the case where $j-\lambda \in \mathbb{Z}$, the representation (\ref{su2_continuous_repr}) is not irreducible since ${J^{+} \, |j\rangle = J^{+} \, |-j-1 \rangle = 0}$, and the corresponding subrepresentation is
    \begin{equation}\label{discrete_subrepresentation}
        \bigl\{ |j-m \rangle : m \in \mathbb{N}_0 \bigl\} \ \cong D^{+}_{j} \ .
    \end{equation}
    In this case the module $C_{j}^{\lambda}$ is reducible but indecomposable, since the complement of (\ref{discrete_subrepresentation}) does not form a subrepresentation. However, the corresponding quotient does:
    \begin{equation}
        C_{j}^{\lambda} \, \big/ \,  \bigl\{ |j-m \rangle : m \in \mathbb{N}_0 \bigl\} \ \cong D^{-}_{j+1} \ .
    \end{equation}
Equivalently, there is a non-split short exact sequence
\begin{equation}\label{continuous_decomp}
    0 \longrightarrow D^{+}_{j} \longrightarrow C^{j}_{j} \longrightarrow D_{j+1}^{-} \longrightarrow 0 ~~~~ \forall \, j \in \mathbb{R} \ .
\end{equation}
\end{itemize}

\subsection{Representations of $\mathfrak{su}(2)_{-1}$}\label{section_affine_representations}

We begin by analyzing the structure of the $\mathfrak{su}(2)_{-1}$ Verma module with highest weight state $|j\rangle $ transforming in $D^{+}_j$ with respect to the zero modes. We prove that there are nontrivial singular vectors only if $j \in \tfrac{1}{2}\mathbb{Z}$, in which case the singular submodule is generated by a single vector. By (\ref{continuous_decomp}) this implies that also the reducible continuous representations with $j \in \tfrac{1}{2}\mathbb{Z}$ and $\lambda = j ~ \text{mod} \,1$ contain a one-dimensional singular submodule. By applying the conjugation automorphism (\ref{conjugation}) one finds the analogous result for a lowest weight affine module.

Before starting the singular vector analysis, we point out that one can elegantly argue that the vacuum Verma module is free of singular vectors using the free field realisation. The vacuum representation of the free fields decomposes in fact into infinitely many $\mathfrak{su}(2)_{-1}$ modules, see (\ref{NS_decomposition}), and this implies the absence of singular vectors in the vacuum Verma module of $\mathfrak{su}(2)_{-1}$ as we argue now. Assume that the latter contains a non-trivial null vector $\mathcal{N}$. Then, by uniqueness of vertex operators, the vertex operator associated to it is zero, and in particular its zero mode $V_{0}(\mathcal{N})$ applied on any highest weight state $|j\rangle $ of the $\mathfrak{su}(2)_{-1}$ theory vanishes:
\begin{equation}
\label{vertex}
    V_{0}(\mathcal{N}) \,|j\rangle  = p(j) \, |j\rangle  = 0 \ ,
\end{equation}
where $p(j)$ is a polynomial in $j$. This follows from the fact that $V_{0}(\mathcal{N})$ applied to $|j\rangle$ has grade zero and hence it can be obtained from $|j\rangle$ by application of $\mathfrak{su}(2)_{-1}$ zero modes, and using the commutator rules (\ref{su2_k_commutators}), it can be expressed as a polynomial in $j$. Since $p(j)$ possesses finitely many roots, the existence of a null vector restricts the set of allowed representations to a finite subset of spins $j$. However, this contradicts (\ref{NS_decomposition}), which is an admissible representation of $\mathfrak{su}(2)_{-1}$ containing infinitely many affine spin $j$ representations. Hence, the vacuum Verma module of $\mathfrak{su}(2)_{-1}$ is free of null vectors.

We now analyze the singular vector structure of all $\mathfrak{su}(2)_{-1}$ Verma modules from first principles. For that, Kac and Kazhdan \cite{Kac} showed that Verma modules can be equipped with an (up to normalization) unique invariant inner product, the \textit{Shapovalov form}, and they gave a formula for its determinant. For the $\mathfrak{su}(2)_k$ Verma module to affine highest weight $(j,k,h_j)$ restricted to the weight space $(j-s,k,h_j+m)$ this takes the form \cite{Ridout:2008nh}  
\begin{equation}
 \begin{aligned}
 \label{determinant}
    \text{det}_{j}(s,m) &:= \prod_{l=1}^{\infty} \bigg\{ \bigl(2j + 1 - l\bigl)^{P(-s + l,\, m)} \prod_{n=1}^{\infty} \bigl(2j +1+ n(k+2) - l \bigl)^{P(-s + l,\,m-nl)} \\
    & \cdot \bigl(-2j -1 + n(k+2) - l \bigl)^{P(-s-l,\,m-nl)} \, \bigl(n(k+2) \bigl)^{P(-s,\,m-nl)} \bigg\} \ ,
\end{aligned}   
\end{equation}
where $P(s,m)$ denotes the multiplicity of the affine weight $(s,0,m)$ in the vacuum Verma module (this is independent of $k$). The presence of a singular vector\footnote{The vanishing of (\ref{determinant}) actually indicates the presence of a subsingular vector, i.e. a vector in the full singular submodule. One is then interested in finding the generators of the singular module, which can be done by the analogous procedure that we outline for $\mathfrak{su}(2)_{-1}$.} in the Verma module translates in the simultaneous vanishing of one of the factors appearing in this formula and the vanishing of the arguments of the function $P$ occurring in the corresponding exponent. Moreover, if a weight is singular, its associated null vector is unique up to normalisation. 

We now specialise to $k =-1$ and therefore omit the $k$-label in the specification of affine weights. Then, (\ref{determinant}) vanishes if and only if one of the following equations holds true:
\begin{equation}
\label{l_equations}
    l =  2j + 1 \ , ~~~~~~l = 2j +1+n  \ , ~~~~~~l = - 2j -1+n  \ .
\end{equation}
Since $l,n \in \mathbb{N}$, the first equation has a solution if and only if $j \in \tfrac{1}{2}\mathbb{N}_0$, and the other two only if $j \in \tfrac{1}{2}\mathbb{Z}$. We distinguish between two cases.

For $j \in \tfrac{1}{2}\mathbb{N}_0$ the first equation always has a solution and the arguments of $P$ in the corresponding exponent vanish for $s = l = 2j +1$ and $m = 0$, so the singular vector has weight $(-j-1 \, ,h_{j} )$, corresponding to the state $(J^{-}_{0})^{2j+1} \, |j \rangle \,$, whose vanishing simply means that the state $|j \rangle$ transforms in the finite-dimensional spin $j$ representations $H_{j}$ of $\mathfrak{su}(2)$. By repeating the same reasoning for the other two equations one finds the singular weights
\begin{equation}
\begin{aligned}
\label{sing_vectors}
     &\bigl(-j-m \,, h_{j}+m  (m-2j-1) \bigl)~~~~\text{for}~~m \geq \max \bigl\{1, 2 (j+1) \bigl\} \ ,\\
     &\bigl(j+m \, , h_{j}+m  (m+2j + 1) \bigl)~~~~~~\,\text{for}~~m \geq \max\bigl\{1, -2j \bigl\} \ .
\end{aligned}
\end{equation}
These seem at first sight to be additional singular vectors, however they all actually belong to the submodule generated by the singular vector ${(-j-1 \, ,h_{j} )}$. Indeed, by repeating the above Kac-Kazhdan analysis for ${(-j-1 \, ,h_{j} )}$ we find exactly the singular weights (\ref{sing_vectors}). As a consistency check, one can also repeat the analysis for all the weights in (\ref{sing_vectors}) and find that the so obtained singular weights are again of the same form, which confirms that they all lie in the same singular submodule. We conclude that the Verma module to highest weight $j$ is irreducible for all $j \in \mathbb{R}\setminus \tfrac{1}{2}\mathbb{Z}$ and for $j \in \tfrac{1}{2}\mathbb{N}_0$ it contains only one (trivial) singular vector at level zero (where by level we refer to the shifted eigenvalue of $L_{0} - h_{j}$), such that the zeroth level states transform in the $H_j$ representation of the zero modes. 

We now consider the case where $j \in -\tfrac{1}{2}\mathbb{N}$. Then, the first equation in (\ref{l_equations}) has no solution, which implies that there is no singular vector at level zero. Note that this is true for all $j \notin \tfrac{1}{2}\mathbb{N}$. The first (lowest level) singular vector obtained from (\ref{sing_vectors}) is $(-j\,, h_{j}- 2j)$, corresponding to the state 
\begin{equation}
\label{sing_vector_D+}
  (J^{+}_{-1})^{-2j} \, |j \rangle \ ,  
\end{equation}
which one can explicitly compute to be singular, i.e. it is a highest weight state of the full affine algebra. Moreover, (\ref{sing_vector_D+}) generates a submodule that contains all the other ones in (\ref{sing_vectors}), as one can confirm by repeating the Kac-Kazhdan analysis for the weight $(-j\,, h_{j}-2j)$. It follows that for $j \in -\tfrac{1}{2}\mathbb{N}$, the quotient of the Verma module of spin $j$ by the submodule generated by the vector (\ref{sing_vector_D+}) is irreducible.

To deduce the analogous result for lowest weight Verma modules, note that the conjugation automorphism induces an isomorphism $D^{-}_{j} \cong (D^{+}_{-j} )^{*}$ which extends at the level of Verma modules. Hence, from above we infer that the singular submodule of the Verma module with affine highest weight states transforming in $D^{-}_{j}$ with $j \in \tfrac{1}{2}\mathbb{N}$ is one-dimensional and generated by 
\begin{equation}\label{sing_vector_D-}
    (J^{-}_{-1})^{2j} \, |j\rangle \ .
\end{equation}

In the following, we denote by $\mathcal{H}_j$ the irreducible $\mathfrak{su}(2)_{-1}$ representation with affine highest weight states transforming in $H_j$ for $j \in \tfrac{1}{2}\mathbb{N}_0$. By the singular vector analysis, $\mathcal{H}_j$ coincides with the Verma module of $H_j$. Analogously, we denote by $\mathcal{D}_j^{\pm}$ the affine representation corresponding to $D^{\pm}_j$, which coincides with the Verma module for ${j \notin \mathbb{R} \setminus  \tfrac{1}{2}\mathbb{Z}}$ and otherwise it is its quotient by the singular submodule generated by (\ref{sing_vector_D+}) and (\ref{sing_vector_D-}) for $j \in  \tfrac{1}{2}\mathbb{Z}$, respectively. Lastly, we denote by $\mathcal{C}^{\lambda}_j$ the affine representation corresponding to $C^{\lambda}_{j}$, which is equal to the Verma module except when $j \in \tfrac{1}{2}\mathbb{Z}$ and $\lambda = j \ \text{mod}\,1$, in which case it is the quotient of the Verma modules by the singular submodule generated by a single vector, as one can deduce by (\ref{continuous_decomp}) and the above analysis\footnote{Note that according to this convention, the modules $\mathcal{D}^{\pm}_{j}$ for $j \in  \pm \tfrac{1}{2}\mathbb{N}_{0}$ and $\mathcal{C}^{\lambda}_j$ for $j \in \tfrac{1}{2}\mathbb{Z}$ and $\lambda = j \ \text{mod}\,1$ are reducible but indecomposable, and while the affine analogous of (\ref{discrete_ses}) holds true, that of (\ref{continuous_decomp}) does not, see (\ref{cont_mod_structure}).}.

Before concluding this section, we spell out the action of the $\mathfrak{su}(2)_{-1}$ automorphisms on the affine modules defined above. For the conjugation automorphism (\ref{conjugation}) we find for every $j$ and $\lambda$ the isomorphisms
\begin{equation}\label{conjugation_isos}
    \bigl(\mathcal{H}_j \bigl)^{*} \cong \mathcal{H}_{j} \ , \qquad \bigl(\mathcal{D}^{\pm}_{j} \bigl)^{*} \cong \mathcal{D}^{\mp}_{-j} \ , \qquad \bigl(\mathcal{C}^{\lambda}_{j} \bigl)^{*} \cong \mathcal{C}^{-\lambda}_{-j}  \ .
\end{equation}
The spectral flow $\sigma$ acts by (\ref{sigma}) with $k = -1$ and generally maps highest weight representations to non-highest weight ones, except for
\begin{equation}
\label{spectral_action}
    \begin{aligned}
        \sigma\bigl(\mathcal{H}_{j}\bigl) &\cong \mathcal{D}^{+}_{-j-\frac{1}{2}} \ ,\qquad \sigma^{-1}\bigl(\mathcal{H}_{j}\bigl) \cong \mathcal{D}^{-}_{j+\frac{1}{2}} \qquad \forall \, j \in \tfrac{1}{2}\mathbb{N}_0 \ ,\\
        \sigma\bigl(\mathcal{D}^{-}_{j}\bigl) &\cong \mathcal{D}^{+}_{j-\frac{1}{2}} \ , \qquad \ \, \sigma^{-1}\bigl(\mathcal{D}^{+}_{j}\bigl) \cong \mathcal{D}^{-}_{j+\frac{1}{2}} \qquad  \forall \, j \notin \tfrac{1}{2} \mathbb{Z} \ .
    \end{aligned}
\end{equation}

\section{Free field realisation}\label{section_free_field_real}

The affine Lie algebra $\mathfrak{u}(2)_{-1}$ has a free field realisation \cite{Goddard:1987td} in terms of two pairs of symplectic bosons $(\lambda^{\alpha}, \mu^{\dagger}_{\alpha})$ with $\alpha = 1,2$, satisfying commutation relations
\begin{equation}\label{free_bos_comm_rel}
    [\lambda^{\alpha}_{r}, (\mu^{\dagger}_{\beta})_{s}] = \delta^{\alpha}_{\beta} \,  \delta_{r,-s} \ .
\end{equation}
We combine these fields into normal ordered bilinears
\begin{equation}
   \tensor{\mathcal{J}}{^\alpha_\beta} =  \mu^{\dagger}_{\beta} \, \lambda^{\alpha} \ , 
\end{equation}
which generate the algebra $\mathfrak{u}(2)_{-1}$. The modes of the generator 
\begin{equation}
    U = \tfrac{1}{2} ( \tensor{\mathcal{J}}{^2_2} + \tensor{\mathcal{J}}{^1_1} )
\end{equation}
are central and form a $\mathfrak{u}(1)_{-1/2}$ algebra, extending the subalgebra $\mathfrak{su}(2)_{-1}$ to $\mathfrak{u}(2)_{-1}$. The other generators are given by
\begin{equation}
    J^{+} = \tensor{\mathcal{J}}{^1_2} \ , \quad J^{-} = \tensor{\mathcal{J}}{^2_1} \ , \quad J^{3} = \tfrac{1}{2} ( \tensor{\mathcal{J}}{^2_2} - \tensor{\mathcal{J}}{^1_1} ) \ ,
\end{equation}
which satisfy the affine $\mathfrak{su}(2)_{-1}$ commutation rules (\ref{su2_k_commutators}) for $k=-1$, with the addition of
\begin{equation}
    [U_{m},U_{n}] = - \tfrac{1}{2}  m  \delta_{m+n,0} \qquad  [U_{m}, J^{a}_{n}] =0 \quad \text{for} \ a = \pm , 3  \ .
\end{equation}
The Sugawara stress-energy tensor is given by 
\begin{equation}
    T^{\mathfrak{u}(2)_{-1}} = T^{\mathfrak{su}(2)_{-1}} + T^{\mathfrak{u}(1)_{-1/2}}= T^{\mathfrak{su}(2)_{-1}} - \, U^{2} \ ,
\end{equation}
where $T^{\mathfrak{su}(2)_{-1}}$ is given by (\ref{Sugawara_T}) with $k=-1$ and normal ordering is assumed.

\subsection{The spectral flow}
\label{section_spec_flow}

As we have seen in eq. (\ref{sigma}) the $\mathfrak{su}(2)_{-1}$ algebra possesses a spectral flow automorphisms, which can be also described directly in terms of the free fields. For this, let us define automorphisms of the symplectic bosons as
\begin{equation}\label{spec_flow_free_fields}
    \sigma^{(\alpha)}\bigl(\lambda_{r}^{\alpha}\bigl) = \lambda^{\alpha}_{r + \frac{1}{2}} \ , \quad \quad  \sigma^{(\alpha)}\bigl((\mu^{\dagger}_{\alpha})_{r}\bigl) = (\mu^{\dagger}_{\alpha})_{r - \frac{1}{2}} \ ,
\end{equation}
for $\alpha = 1,2$. The combination 
\begin{equation}\label{sigma_2}
    \sigma = \sigma^{(1)} \circ \left(\sigma^{(2)} \right)^{-1} \ ,
\end{equation}
leaves the $\mathfrak{u}(1)_{-1/2}$ algebra invariant and acts on $\mathfrak{su}(2)_{-1}$ as (\ref{sigma}) for $k=-1$ (so by slight abuse of notation we identify the two).
The other natural combination is
\begin{equation}
    \sigma_{U} = \bigl(\sigma^{(1)}\bigl)^{-1} \circ \bigl(\sigma^{(2)}\bigl)^{-1} \ ,
\end{equation}
which leaves the subalgebra $\mathfrak{su}(2)_{-1} $ invariant and on the other generators it acts as
\begin{equation}\label{sigmaU_action}
        \sigma^{w}_{U}(U_{m}) = U_{m} - \tfrac{w}{2}  \delta_{m,0} \qquad \text{and} \qquad \sigma^{w}_{U}(L_{0}^{\mathfrak{u}(2)_{-1}}) = L_{0}^{\mathfrak{u}(2)_{-1}} + w  U_{0} - \tfrac{w^{2}}{4} \ . 
\end{equation}

\subsection{Free field representations}
\label{section_free_field_representations}

Next we want to discuss the representation arising from the free fields. As usual, in the NS sector the free fields are half-integer moded and the full affine representation is generated by the action of the negative modes on a single highest weight state~$|0\rangle$ which is annihilated by all zero modes. We denote this affine representation by $\mathcal{V}$, which is the vacuum representation of the symplectic bosons, and by $\mathcal{V}_U$ the subsector with fixed $U_0=U \in \tfrac{1}{2}\mathbb{Z}$ eigenvalue. In particular, $\mathcal{V}_0$ contains the $\mathfrak{su}(2)_{-1}$ representation~$\mathcal{H}_0$. 

On the other hand, in the R sector the free fields are integer-moded and we label the highest weight states by the symplectic boson occupation numbers $|m_1, m_2 \rangle$. We then define the action of the symplectic boson zero modes  by\footnote{\label{footn}There is freedom in defining the action of the symplectic boson zero modes, reflected by the \textit{conjugation} automorphism of the free fields 
\begin{equation}\label{conj_aut}
    \bigl(\lambda_r^{\alpha} \bigl)^{*} = (\mu^{\dagger}_{\alpha})_r \quad \text{and} \quad \bigl((\mu_{\alpha}^{\dagger})_r \bigl)^{*} = - \lambda^{\alpha}_r \quad \text{for} \ \alpha = 1,2 \ .
\end{equation}
There are therefore four different R sectors, whose distinction is only relevant when considering the subsectors $m_1,m_2 \in \tfrac{1}{2}\mathbb{N}_0$, which yield different $\mathfrak{su}(2)_{-1}$ representations. We discuss this distinction in Appendix \ref{app_different_R}.} 
\begin{equation}\label{sympl_bos_zero}
     \begin{aligned}[t]
     \lambda^{1}_{0} \, |m_{1}, m_{2} \rangle & = 2  m_{1} \, | m_{1} - \tfrac{1}{2}, m_{2} \rangle \ , \\
    \lambda^{2}_{0} \, |m_{1}, m_{2} \rangle & = 2  m_{2} \, | m_{1} , m_{2} - \tfrac{1}{2} \rangle \ ,
    \end{aligned}
    \qquad \qquad
    \begin{aligned}[t]
    (\mu^{\dagger}_{1})_{0} \, |m_{1}, m_{2} \rangle & = | m_{1} + \tfrac{1}{2}, m_{2} \rangle \ , \\
    (\mu^{\dagger}_{2})_{0} \, |m_{1}, m_{2} \rangle & = | m_{1} , m_{2} + \tfrac{1}{2} \rangle \ ,
    \end{aligned}
\end{equation}
where $m_{i} \in \tfrac{1}{2}\mathbb{Z} + \delta_{i}$ with $\delta_{1}, \delta_{2} \in \mathbb{R}/\tfrac{1}{2}\mathbb{Z} \cong [0,\tfrac{1}{2})$. 
With this convention, one computes the action of the $\mathfrak{u}(2)_{-1}$ generators\footnote{We use the usual normal ordering convention that positive modes stand to the right of negative modes. Furthermore, we define $: a_{0}b_{0} : \, = \tfrac{1}{2} ( a_{0}b_{0} + b_{0} a_{0} )$ for bosonic zero modes $a_0,b_0$.}
\begin{equation}\label{J_U_action}
    \begin{aligned}
     J^{3}_{0} \, | m_{1}, m_{2} \rangle &= (m_{2}   - m_{1}  ) \, | m_{1}, m_{2} \rangle \ , \\
     J^{+}_{0} \, | m_{1}, m_{2} \rangle &= 2  m_{1} \,  | m_{1} - \tfrac{1}{2}, m_{2} + \tfrac{1}{2} \rangle \ , \\
     J^{-}_{0} \, | m_{1}, m_{2} \rangle &= 2  m_{2} \, | m_{1} + \tfrac{1}{2}, m_{2} - \tfrac{1}{2} \rangle \ , \\
     U_{0} \, | m_{1}, m_{2} \rangle &= (m_{1}   + m_{2} + \tfrac{1}{2}) \, | m_{1}, m_{2} \rangle \ , \\
    \end{aligned}
\end{equation}
as well as the $\mathfrak{su}(2)_{-1}$ Casimir 
\begin{equation}
    C^{\mathfrak{su}(2)} = J^{3}_{0}  J^{3}_{0} + \tfrac{1}{2}  (J^{+}_{0}  J^{-}_{0} + J^{-}_{0} J^{+}_{0}) = j  (j + 1)
\end{equation}
with $j = m_{1} + m_{2} $. The full affine R sector is then generated by the action of the free field negative modes on the highest weight states $|m_{1},m_{2} \rangle$. We denote by $\mathcal{R}^{\lambda}_{U}$ the subrepresentation at fixed $U_{0} = U \in \mathbb{R}$ of the R sector defined by (\ref{sympl_bos_zero}) with 
\begin{equation}\label{lambda}
\begin{aligned}
        \delta_{1} = \frac{U-\lambda}{2} \ \text{mod} \, \tfrac{1}{2} \quad \text{and} \quad
           \delta_{2} = \frac{U+\lambda}{2} \ \text{mod} \, \tfrac{1}{2} \ .
\end{aligned}
\end{equation}
It is straightforward to check that the highest weight states of $\mathcal{R}^{\lambda}_{U}$ transform in the $\mathfrak{su}(2)$ continuous representation $C_{U-1/2}^{\lambda -1/2}$, which is irreducible except when $\lambda = U \,\text{mod}\, 1$ in which case it is reducible but indecomposable. 

In the following we denote by $\mathcal{R}$ the R sector with
\begin{equation}\label{special_deltas}
   \delta_{1} = \delta_{2} = 0 \quad \iff \quad \lambda = U \ \text{mod}\,1 \,, \ U \in \tfrac{1}{2}\mathbb{Z} 
\end{equation}
and by $\mathcal{R}^+$ its subrepresentation defined by $m_{1},m_{2} \in \tfrac{1}{2}\mathbb{N}_0$. We label by a subscript $U$ the corresponding subsectors to fixed $U_0=U \in \tfrac{1}{2}\mathbb{Z}$. Note that $\mathcal{R}_U=\mathcal{R}_U^U$.

Free field spectral flows (\ref{spec_flow_free_fields}) switch between integer and half-integer moding, i.e. they interchange NS and R sectors. In fact, there is an isomorphism
 \begin{equation}\label{sigmaU_iso}
        \sigma_{U}^{-1}\bigl(\mathcal{V}\bigl) \cong \mathcal{R}^{+} \ ,
\end{equation}
following directly from the identification $|0,0\rangle = \bigl[|0\rangle \bigl]^{\sigma_U^{-1}}$, since $\bigl[|0\rangle \bigl]^{\sigma_U^{-1}}$ is annihilated by $\lambda_0^{\alpha}$ but not by $(\mu^{\dagger}_\alpha)_0$.

\section{Characters}
\label{section_characters}

Now that we analyzed the structure of the irreducible affine $\mathfrak{su}(2)_{-1}$ modules, the next step is to compute their characters. We do this for both the affine $\mathfrak{su}(2)_{-1}$ and for the free field modules in Section \ref{section_free_field_characters}.

We start from the $\mathfrak{su}(2)_{-1}$ characters. We write as usual $q = e^{2 \pi i \tau}$ for $\tau$ in the complex upper half-plane corresponding to the chemical potential of $L_0$, and $x = e^{2 \pi i t}$ for $t \in \mathbb{C}$ corresponding to the chemical potential of $J^{3}_0$. In our context, a \textit{character} $\text{ch}\bigl[\mathcal{M}\bigl](t;\tau)$ of an affine module $\mathcal{M}$ is a formal distribution, or equivalently, a formal power series. Some characters can be also expressed as meromorphic functions of $x$ and $q$, expanded on a specific convergence domain in the $x$-plane (an annulus depending on $|q|$). We call such meromorphic extension a \textit{character function}, and we will always specify the convergence domain on which it agrees with a certain character. 

An important observation \cite{Ridout:2008nh} is that the association of isomorphism classes of irreducible modules to characters is one-to-one, while the association to character functions is not injective. Differently phrased, non-isomorphic modules can possess the same character function, and their actual character is obtained by expanding the character function on disjoint convergence regions. This feature is characteristic of non-unitary models in which the spectral flow automorphisms have infinite orbit \cite{Creutzig:2012sd,Creutzig:2013yca,Lesage:2003kn}.

\subsection{$\mathfrak{su}(2)_{-1}$ characters}

By the null vector analysis, for $j \in \tfrac{1}{2}\mathbb{N}$ we have
\begin{equation}\label{character_j_finite_dim}   \text{ch}\bigl[\mathcal{H}_{j}\bigl](t;\tau)  = \text{ch}\bigl[\mathcal{H}_{0}\bigl](t;\tau) \,  q^{j(j+1)}  \chi_{j}^{\mathfrak{su}(2)}(t)  =  \frac{i    q^{(j+\frac{1}{2})^2}  \bigl(x^{j+\frac{1}{2}} - x^{-j-\frac{1}{2}} \bigl)}{\vartheta_{1}(t;\tau)}  \ ,
\end{equation}
where the second equality is valid only on the convergence region $|q|<|x|<|q|^{-1}$, $\chi^{\mathfrak{su}(2)}_{j}(t)$ is the character of the spin $j$ representation of $\mathfrak{su}(2)$
\begin{equation}
        \chi^{\mathfrak{su}(2)}_{j}(t) = \sum_{m=-j, \, m+j \,\in \mathbb{Z}}^{j} x^{m} = \frac{x^{j + \frac{1}{2}} - x^{-j-\frac{1}{2}}}{x^{\frac{1}{2}} - x^{-\frac{1}{2}}}  \ ,
\end{equation}
and  
\begin{equation}  \text{ch}\bigl[\mathcal{H}_{0}\bigl](t;\tau)   = q^{\frac{1}{8}}  \frac{1}{\prod_{n\geq 1}(1-x  q^{n})  (1-q^{n})  (1-x^{-1}  q^{n})} 
\end{equation}
is the character of the irreducible vacuum Verma module again for $|q|<|x|<|q|^{-1}$.
Note that $\vartheta_{1}(t;\tau)$ has zeros at $x=q^{n}$ for all $n\in \mathbb{Z}$ and the vanishing of the denominator in (\ref{character_j_finite_dim}) at $x = 1$ is compensated by the vanishing of the numerator. Hence, the convergence region can be extended from $1<|x|<|q|^{-1}$ to $|q|<|x|<|q|^{-1}$, feature which is peculiar of characters of affine modules generated by finite-dimensional $\mathfrak{su}(2)$ representations. 

For the discrete highest weight affine representations of spin $j \in \mathbb{R} \setminus \tfrac{1}{2}\mathbb{Z}$ we have
\begin{equation}
\label{character_j+}   \text{ch}\bigl[\mathcal{D}^{+}_{j}\bigl](t;\tau)  = \text{ch}\bigl[\mathcal{H}_{0}\bigl](t;\tau) \, q^{j(j+1)} \sum_{m \leq j}x^{m}   = \frac{i  q^{(j + \frac{1}{2})^{2}}  x^{j+\frac{1}{2}}}{\vartheta_{1}(t;\tau)} =: \chi^{+}_{j}(t;\tau) \ ,
\end{equation}
valid for $1<|x|<|q|^{-1}$. The character of its conjugate representation is 
\begin{equation}\label{character_j-} 
\text{ch}\bigl[\mathcal{D}^{-}_{j}\bigl](t;\tau)  = \text{ch}\bigl[\mathcal{H}_{0}\bigl](t;\tau) \, q^{j(j-1)} \sum_{m \geq j}x^{m} = -\frac{i  q^{(j - \frac{1}{2})^{2}}  x^{j-\frac{1}{2}}}{\vartheta_{1}(t;\tau)} =: \chi^{-}_{j}(t;\tau) \ ,
\end{equation}
where the second equality holds for $|q|<|x|<1$. We point out that there is an identity of formal power series
\begin{equation}\label{D+_D__relation}
\text{ch}\bigl[\mathcal{D}^{+}_{-j}\bigl](t;\tau) = \text{ch}\bigl[\mathcal{D}^{-}_{j}\bigl](-t;\tau) \qquad \forall \, j \in \mathbb{R} \setminus \tfrac{1}{2}\mathbb{Z} \ ,
\end{equation}
reflecting (\ref{conjugation_isos}).
By the analysis of Section \ref{section_affine_representations} we compute
\begin{equation}\label{su_1_char_discrete1}   \text{ch}\bigl[\mathcal{D}^{\pm }_{j} \bigl](t;\tau) = \chi^{\pm}_{j}(t;\tau) \, \bigl(1-q^{\mp 2j}  x^{-2j} \bigl) \qquad \forall \, j \in \mp \tfrac{1}{2}\mathbb{N} \ ,
\end{equation}
which yields the affine version of (\ref{discrete_ses})
\begin{equation}
    \text{ch}\bigl[\mathcal{D}^{\pm}_{j} \bigl] = \text{ch}\bigl[\mathcal{H}_{j} \bigl] + \, \text{ch}\bigl[\mathcal{D}^{+}_{\pm(j+1)} \bigl] \qquad \forall \, j \in \pm \tfrac{1}{2}\mathbb{N}_0  \ .
\end{equation}
Note that since in (\ref{su_1_char_discrete1}) the singularity at $x = q^{\mp 1}$ is removable, the convergence region can be extended, e.g. from $1<|x|<|q|^{-1}$ to $1<|x|<|q|^{-2}$ for $\mathcal{D}^+_{j}$ and similarly for~$\mathcal{D}^-_{j}$. 

In Section \ref{section_affine_representations} we found that the Verma module associated to $C^{\lambda}_j$ is free of singular vectors except when $\lambda = j = 0,\tfrac{1}{2} ~\text{mod}\,1$. In the former case, its affine character is given by 
\begin{equation}\label{cont_su2__1_char}   \text{ch}\bigl[\mathcal{C}_{j}^{\lambda}\bigl](t;\tau) = \text{ch}\bigl[\mathcal{H}_{0}\bigl](t;\tau) \, q^{j(j+1)} \sum_{m \in \mathbb{Z}}x^{\lambda+m} = \frac{ q^{(j + \frac{1}{2})^{2}}   \sum_{m \in \mathbb{Z}} x^{\lambda + m}}{\eta(\tau)^{3}}   =: \chi^{\lambda}_{j}(t;\tau) \ ,
\end{equation}
where we used the identity
\begin{equation}
    \frac{\sum_{m \in \mathbb{Z}}x^{m}}{\prod_{n\geq 1}(1-x  q^{n})  (1-x^{-1}  q^{n})} = \frac{ q^{\frac{1}{12}} \sum_{m \in \mathbb{Z}} x^{ m}}{\eta(\tau)^{2}}  \ .
\end{equation}
For $j \in \tfrac{1}{2}\mathbb{Z}$ and $\lambda = j ~\text{mod} \, 1$, the situation is quite different. Since $j$ and $-j-1$ parameterise the same module, we restrict our attention to $j \geq -1/2$. Then the module $C^{j}_{j}$ is indecomposable and its structure is given by eq. (\ref{continuous_decomp}) but this does not translate directly to the corresponding affine modules. As we have seen, for $j \in \tfrac{1}{2}\mathbb{Z}$ and $j \geq -\tfrac{1}{2}$ the Verma module of $C^{j}_{j}$ has a unique singular vector, so its irreducible quotient has character
\begin{equation}\label{cont_char_special}               \text{ch}\bigl[\mathcal{C}^{j}_{j} \bigl](t;\tau) = \text{ch}\bigl[\mathcal{C}^{j}_{-j-1} \bigl](t;\tau) 
        = \chi^{j}_{j}(t;\tau) \, \bigl(1-q^{2(j+1)}x^{2(j+1)} \bigl) 
        = \chi^{j}_{j}(t;\tau) - \chi^{j}_{j+1}(t;\tau) \ ,
\end{equation}
for $j \in \tfrac{1}{2}\mathbb{Z}$.
Therefore, we find following identities
\begin{equation}\label{cont_mod_structure}
    \text{ch}\bigl[\mathcal{C}^{j}_{j} \bigl]
    = \begin{cases}
            \text{ch}\bigl[\mathcal{D}^{+}_{-1/2} \bigl] + \, \text{ch}\bigl[\mathcal{H}_{1/2}\bigl] + \, \text{ch}\bigl[\mathcal{D}^{-}_{1/2} \bigl]  ~~~~~~~~~~~~~~~~~\, \text{if}~~ j=-1/2 \ , \\
            \text{ch}\bigl[\mathcal{D}^{+}_{-j-1} \bigl] + \, \text{ch}\bigl[\mathcal{H}_{j}\bigl] + \, \text{ch}\bigl[\mathcal{D}^{-}_{j+1} \bigl] + \, \text{ch}\bigl[\mathcal{H}_{j+1}\bigl] ~~~~ \text{if}~~ j \geq 0 \ ,
        \end{cases}
\end{equation}
which give the affine version of (\ref{continuous_decomp}).

Let us consider the spectrally flowed affine modules.
By (\ref{sigma}) with $k=-1$ and (\ref{spec_flo_action}), any $\mathfrak{su}(2)_{-1}$ character transforms as
\begin{equation}\label{sigma_action_modules}
    \text{ch} \bigl[ \sigma^{w}\bigl(\mathcal{M} \bigl) \bigl](t;\tau) = x^{\frac{-w}{2}} \,  q^{\frac{w^2}{4}} \text{ch} \bigl[ \mathcal{M}  \bigl] (t - w \tau;\tau) \quad \forall \, w \in \mathbb{Z} \ .
\end{equation}
It is important to point out that $\sigma^{w}$ shifts the convergence region of character functions in the $x$-plane by a factor $|q|^{-w}$. For instance, using (\ref{theta_function_periodicity1}) one computes 
\begin{equation}
\label{tilde_char_specflow}  
\text{ch} \bigl[\sigma^{w}\bigl(\mathcal{D}^{\pm}_{j}\bigl) \bigl](t;\tau) =  (-1)^{w} \bigl[\sigma^{w}\bigl(\mathcal{D}^{\pm}_{j+\frac{w}{2}}\bigl) \bigl](t;\tau)    ~~~~ \forall \, j \in \mathbb{R} \setminus \tfrac{1}{2}\mathbb{Z} \ ,
\end{equation}
where the equality in terms of meromorphic functions holds on the convergence region $|q|^{-w} < |x|<|q|^{-w-1}$ for $+$ and $|q|^{-w +1} < |x|<|q|^{-w}$ for $-$. From this we can see that the association of (isomorphisms classes of) irreducible $\mathfrak{su}(2)_{-1}$ modules to their character functions is not injective.

\subsection{Free field characters }
\label{section_free_field_characters}

We now compute the characters of the free field representations and decompose them in terms of $\mathfrak{su}(2)_{1}$ characters. This computation can be interestingly also obtained from a particular denominator identity for Lie superalgebras found by Kac \cite{Kac:1994kn} and answers one of the questions raised in \cite{Gaberdiel_2018}, which was indeed about the structure of the free field representations of four symplectic in $\mathfrak{su}(2)_{1}$ modules. 

Four real (or two complex) fermions realise the affine algebra $\mathfrak{so}(4)_{1} \cong \mathfrak{su}(2)_{1} \oplus \mathfrak{su}(2)_{1}$. The situation is quite different for four symplectic bosons, since it is not possible to extend the $\mathfrak{u}(1)$ generator $U_0$ to another commuting $\mathfrak{su}(2)$ algebra as in the case of fermions. More precisely, the bilinears formed by two pairs of symplectic bosons generate $\mathfrak{sp}(4)_{-1/2}$, which contains ${\mathfrak{u}(2)_{-1} = \mathfrak{su}(2)_{-1} \oplus \mathfrak{u}(1)_{-1/2}}$ as the $U_{0}$-uncharged subalgebra, but it is bigger since at the zero mode level  
\begin{equation}
\label{sp4_decomposition}
    \mathfrak{sp}(4) \cong \mathfrak{su}(2) \oplus \mathfrak{u}(1) \oplus \mathbf{3}_{1} \oplus \overline{\mathbf{3}}_{-1} \ ,
\end{equation}
where $\mathbf{3}_{1}$ and $\overline{\mathbf{3}}_{-1}$ denote the $3$-dimensional representation of $\mathfrak{su}(2)$ and its conjugate, respectively, and the subscript labels the $U_{0}$-eigenvalue. The two adjoint representations are generated by the two pairs of three independent bilinears $\lambda^{\alpha} \, \lambda^{\beta}$ and $\mu^{\dagger}_{\alpha} \, \mu^{\dagger}_{\beta}$ respectively. The latter generate an ideal of $\mathfrak{sp}(4)$ but not a subalgebra, and hence there is no decomposition analogous to the fermionic case. The relevant ($U_{0}$-uncharged) algebra associated to two pairs of symplectic bosons is then $\mathfrak{u}(2)_{-1}$, which has central charge $c =  -2$, in agreement with the central charge of the four symplectic bosons\footnote{This is of course the same central charge as $\mathfrak{sp}(4)_{-1/2}$.}, each contributing $c = - \tfrac{1}{2}$. Because of (\ref{sp4_decomposition}), we expect the $U_{0}$-fixed part of the free field representations to be reducible as $\mathfrak{su}(2)_{-1}$ modules, which is indeed the case for the NS sector and for the R sector with special (non-generic) discrete values of parameters as in (\ref{special_deltas}), as we show now.

We start by the R sector defined by (\ref{sympl_bos_zero}) and (\ref{lambda}), whose character is
\begin{equation}
\begin{aligned}\label{char_sympl_bos}
& q^{-\frac{1}{6}}\sum_{m_1 \in  \frac{1}{2} \mathbb{Z} + \delta_1} \ \sum_{m_2 \in \frac{1}{2} \mathbb{Z} + \delta_2} \, x^{m_2-m_1 } \, y^{m_1 + m_2 +\frac{1}{2}} \, \prod_{n=1}^{\infty} \prod_{a,b = \pm \frac{1}{2}} \frac{1}{1- x^a y^b q^n} \   \\ 
& = \left( \sum_{r \in \mathbb{Z} + \lambda} \, \sum_{s \in \mathbb{Z} + U+ \frac{1}{2} } + \sum_{r \in  \mathbb{Z}+\lambda + \frac{1}{2}} \, \sum_{s \in \mathbb{Z} + U} \right)\, 
x^r\, y^s \, \frac{1}{\eta(\tau)^4}  \ ,
\end{aligned}
\end{equation}
where we denote by $y = e^{2 \pi i \mu}$ the chemical potential associated to~$U_{0}$. Note that (\ref{char_sympl_bos}) converges nowhere in the $(x,y)$-plane and must thus be treated as a formal distribution, see discussion at the beginning of Section \ref{section_characters}. We can thus write
\begin{equation}\label{R_U_lambda_char}  \text{ch}\bigl[\mathcal{R}_U^{\lambda} \bigl](t;\tau) = \sum_{r \in \mathbb{Z} + \lambda + \frac{1}{2} }x^{r} \, \frac{1}{\eta(\tau)^{4}} = \frac{q^{-U^2}}{\eta(\tau)} \, \chi^{\lambda-\frac{1}{2}}_{U-\frac{1}{2}}(t;\tau) \ ,
\end{equation}
and for generic values of $\lambda$ and $U$ (or equivalently of $\delta_1$ and $\delta_2$), by comparing (\ref{R_U_lambda_char}) with (\ref{cont_su2__1_char}) it follows that $\mathcal{R}^{\lambda}_U$ is an irreducible $\mathfrak{u}(2)_{-1}$ module
\begin{equation} \mathcal{R}^{\lambda}_{U} \cong \mathcal{F}_{U} \otimes \mathcal{C}^{\lambda-\frac{1}{2}}_{U-\frac{1}{2}}  \quad \text{unless $U \in \tfrac{1}{2}\mathbb{Z}$ and $\lambda = U \ \text{mod} \, 1$} \ ,
\end{equation}
where $\mathcal{F}_{U}$ denotes the $\mathfrak{u}(1)_{-1/2}$ irreducible module with highest weight $U$, which is simply the Fock space of a single boson.
On the other hand, for the special values of $U$ and $\lambda$, the R sector $\mathcal{R}$ has a non-trivial decomposition. Indeed, combining (\ref{R_U_lambda_char}) with (\ref{cont_char_special}) we find
\begin{equation}\label{R_decomp}
    \mathcal{R} \cong \bigoplus_{U \in \frac{1}{2}\mathbb{Z}}  \mathcal{R}_{U} \quad \text{with} \quad \mathcal{R}_{ U} \cong \mathcal{F}_{U} \otimes \bigoplus_{j \in \mathbb{N} + |U|-\frac{1}{2}} \mathcal{C}^{j}_{j}  \ .
\end{equation}
We can transfer (\ref{R_decomp}) to the subsector $\mathcal{R}^{+} \subset \mathcal{R}$,
\begin{equation}
\label{R_decomposition}
    \mathcal{R}^{+} \cong \bigoplus_{U \in \frac{1}{2}\mathbb{Z}} \mathcal{R}^{+}_{U} \quad \text{with} \quad \mathcal{R}^{+}_{U} \cong \mathcal{F}_{U} \otimes \bigoplus_{j\in \mathbb{N}+ |U-\frac{1}{2}|}  \mathcal{H}_{j} \ ,
\end{equation}
whose full character is
\begin{equation}
\label{R_character}
    \begin{aligned}
        \text{ch} \bigl[\mathcal{R}^+\bigl](t,\mu;\tau) &= q^{-\frac{1}{6}} \sum_{m_{1},m_{2} \in \frac{1}{2}\mathbb{N}} x^{m_{2}-m_{1}} \, y^{m_1+m_{2} + \frac{1}{2}} \prod_{n=1}^{\infty} \prod_{a,b = \pm \frac{1}{2}} \frac{1}{1 - x^{a} y^{b}  q^{n} } \\       
        &= \frac{\eta(\tau)^2}{\vartheta_{1}(\frac{t+\mu}{2};\tau)\vartheta_{1}(\frac{t-\mu}{2};\tau)}  \ ,
    \end{aligned}
\end{equation}
valid for $|q|<|x|<|q|^{-1}$ and $|q|^2 < |y|<1$. Then (\ref{R_decomposition}) translates to 
\begin{equation}\label{R_character_decomposition}
     \text{ch} \bigl[\mathcal{R}^{+}_U\bigl](t;\tau) = \frac{q^{-U^2}}{\eta(\tau)} \sum_{j \in \mathbb{N}+|U-\frac{1}{2}|} \text{ch}\bigl[\mathcal{H}_j \bigl](t;\tau)   \ .
\end{equation}
We argue that
(\ref{R_character_decomposition}) follows independently from a particular denominator identity for Lie superalgebras \cite{Kac:1994kn}, which states that for $u,v \in \mathbb{C}$ with $|q|<|u|\,,|v|<1$ we have 
\begin{equation}
\label{denominator_identity}
    \prod_{n=1}^{\infty} \frac{(1-q^{n})^2(1-uvq^{n-1})(1-u^{-1}v^{-1}q^{n})}{(1-uq^{n-1})(1-u^{-1}q^{n})(1-vq^{n-1})(1-v^{-1}q^{n})} = \Bigg(\sum_{m,n=0}^{\infty}-\sum_{m,n=-1}^{-\infty} \Bigg)u^{m}v^{n}q^{mn} \ .
\end{equation}
By substituting $u = x^{\frac{1}{2}}y^{-\frac{1}{2}}$, $v = x^{\frac{1}{2}}y^{\frac{1}{2}}$ we obtain
\begin{equation}
\begin{aligned}
        \sum_{m,n=0}^{\infty}x^{\frac{n-m}{2}}y^{\frac{n+m}{2}} \prod_{n=1}^{\infty} \frac{(1-q^{n})^2(1- x q^{n})(1-x^{-1} q^{n})}{(1- x^{\frac{1}{2}}y^{-\frac{1}{2}}q^{n})(1-x^{-\frac{1}{2}}y^{\frac{1}{2}}q^{n})(1-x^{\frac{1}{2}}y^{\frac{1}{2}} q^{n})(1- x^{-\frac{1}{2}}y^{-\frac{1}{2}}q^{n})}  \\
    = \Bigg(\sum_{m,n=0}^{\infty}-\sum_{m,n=-1}^{-\infty} \Bigg)\frac{x^{\frac{n+m+1}{2}}}{x-1}y^{\frac{m-n-1}{2}}q^{mn} \ ,
\end{aligned}
\end{equation}
which proves (\ref{R_character_decomposition}). 

For what concerns the NS sector we use (\ref{sigmaU_iso}) and obtain
\begin{equation}
\label{NS_decomposition}
    \mathcal{V} = \bigoplus_{U \in \frac{1}{2}\mathbb{Z}} \mathcal{V}_{U} \quad \text{with} \quad \mathcal{V}_{U} \cong \mathcal{F}_{U} \otimes \bigoplus_{j \in \mathbb{N} + |U|} \mathcal{H}_{j} \ ,
\end{equation}
whose character is
\begin{equation}
\label{NS_level_1}
\begin{aligned}
     \text{ch} \bigl[\mathcal{V}\bigl](t,\mu;\tau) &=  q^{\frac{1}{12}}  \prod_{n = 1}^{\infty} \prod_{a,b = \pm \frac{1}{2}} \frac{1}{1 - x^{a} y^{b}  q^{n - \frac{1}{2}} } = \frac{\eta(\tau)^2}{\vartheta_{4}(\frac{t+\mu}{2};\tau)  \vartheta_{4}(\frac{t-\mu}{2};\tau)}  \ ,
\end{aligned}
\end{equation}
valid for $|q|<|x|,|y|<|q|^{-1}$. Then (\ref{NS_decomposition}) translates to
\begin{equation}\label{NS_dec_char}
     \text{ch} \bigl[\mathcal{V}_U\bigl](t;\tau)  = \frac{q^{-U^2}}{\eta(\tau)} \sum_{j\in \mathbb{N} + |U|} \text{ch}\bigl[\mathcal{H}_j \bigl](t;\tau)   \ ,
\end{equation}
which can also be proven by substituting $u = q^{\frac{1}{2}}x^{\frac{1}{2}}y^{-\frac{1}{2}}$, $v = q^{\frac{1}{2}}x^{\frac{1}{2}}y^{\frac{1}{2}}$ in (\ref{denominator_identity}), yielding
\begin{equation}
 \left[ \prod_{n = 1}^{\infty} \prod_{a,b = \pm \frac{1}{2}} \frac{1}{1 - x^{a}\, y^{b} \, q^{n - \frac{1}{2}} } \right]_{\mu=U} = q^{U} \left[  \sum_{s \in \frac{1}{2}\mathbb{N}} y^{s} \sum_{m=-s}^{s} x^{m} \prod_{n=1}^{\infty} \prod_{a,b = \pm \frac{1}{2}} \frac{1}{1 - x^{a}\, y^{b} \, q^{n} } \right]_{\mu = U} \ ,
\end{equation}
for every $U \in \frac{1}{2}\mathbb{Z}$.

\section{Modular invariants}
\label{section_mod_inv}

In order to have a well-defined string theory at one-loop, the WZW model spectrum must be invariant under the modular group ${\rm SL}(2,\mathbb{Z})$, the symmetry group of the moduli space of complex structures of the torus. This translates into the requirement that the WZW model partition function must be modular invariant. We therefore investigate the modular transformation properties of the $\mathfrak{su}(2)_{-1}$ characters. We find that the characters of the affine modules carry (an infinite dimensional) representation of ${\rm SL}(2,\mathbb{Z})$. This allows us to find certain continuous and discrete invariant spectra for the $\mathfrak{su}(2)_{-1}$ WZW model.

\subsection{Finite-dimensional and discrete representations}
\label{section_fin_dim_discr}

We can compute the modular transformations of the discrete character functions of eqs. (\ref{character_j+}) and (\ref{character_j-}) as
\begin{equation}\label{su2__1_S_T_matrices}
\begin{aligned}
     \chi^{\pm}_{j}(t;\tau+1) &= e^{2\pi i \left[j(j\pm1)+ \frac{1}{8} \right]}  \chi^{\pm}_{j}(t;\tau) \ ,\\  
     \chi^{\pm}_{j}\bigl(\tfrac{t}{\tau}; -\tfrac{1}{\tau}\bigl) &= e^{-\frac{\pi i t^{2}}{2 \tau}}  \int_{\mathbb{R}} dj' \,  \sqrt{2}  i  e^{-\pi i (2j \pm 1)(2j' \pm 1)} \chi^{\pm}_{j'}(t;\tau) \ ,
\end{aligned}
\end{equation}
so that the modular $T$- and $S$-matrix are given by
\begin{equation}\label{S_T_matrices}
    T^{\pm }_{jj'} = e^{2\pi i \left[ j(j\pm1)+ \frac{1}{8} \right]}   \delta(j-j')~~~~\text{and}~~~~S^{\pm}_{jj'} = \sqrt{2}  i   e^{-\pi i (2j \pm 1)(2j' \pm 1)} \ ,
\end{equation}
respectively. Note that the finite-dimensional character functions (\ref{character_j_finite_dim}) transform in the same way as $\chi^{+}_{j}$. The modular matrices (\ref{S_T_matrices}) are formally symmetric and unitary, in the sense that (dropping the superscripts)
\begin{equation}
    \int_{\mathbb{R}} dj' \, T_{jj'}  T_{j'j''}^{\dagger} =  \int_{\mathbb{R}} dj' \, S_{jj'}  S_{j'j''}^{\dagger} = \delta(j-j'') \ .
\end{equation}
Therefore, the diagonal partition functions 
\begin{equation}\label{part_funct_discr}
    \int_{\mathbb{R}} dj \, \chi^{\pm}_{j}(t;\tau)  \, \overline{\chi^{\pm}_{j}(t;\tau)} = \frac{1}{2}  \frac{1}{\sqrt{\Ima(\tau)}}  \frac{e^{2\pi \frac{\Ima(t)^{2}}{\Ima(\tau)}}}{\bigl|\vartheta_{1}(t;\tau) \bigl|^{2}} \,
\end{equation}
are modular invariant. Note that (\ref{part_funct_discr})
agrees with the contribution of the discrete representations $\mathcal{D}^{+}_{j}$ to the partition function of the ${\rm SL}(2,\mathbb{R})$ WZW model at level $k>2$, whose allowed (unitary) range of spins is $1/2<j<(k-1)/2$ \cite{Maldacena:2000hw}. This is to be expected since all the computations of \cite{Maldacena:2000hw} work out in the same way by inserting $k=1$, and in particular the discrete representations $\mathcal{D}^{+}_{j}$ of $\mathfrak{su}(2)_{-1}$ on the range $-1/2<j<0$ combine with the spectral flow sum to give the range $j \in \mathbb{R} \setminus \tfrac{1}{2}\mathbb{Z}$ in the integral (\ref{part_funct_discr}) at the level of character functions, see (\ref{tilde_char_specflow}). We can thus write the spectra corresponding to (\ref{part_funct_discr}) as
\begin{equation}\label{discr_spectra}
    \mathcal{H}^{\mathfrak{su}(2)_{-1}}_{\rm discr, \pm } = \pm \OplusInt_{(\mp\frac{1}{2},{0})} dj \bigoplus_{w \in \mathbb{Z}} \, \sigma^{w}\bigl(\mathcal{D}^{\pm}_{j}\bigl) \otimes \, \overline{\sigma^{w}\bigl(\mathcal{D}^{\pm}_{j}\bigl)} \ ,
\end{equation}
which are modular invariant spectra of $\mathfrak{su}(2)_{-1}$, or equivalently of $\mathfrak{sl}(2,\mathbb{R})_1$.
Note that the spectra (\ref{discr_spectra}) have the same partition function (\ref{part_funct_discr}) but are not conjugation invariant:
\begin{equation}
    \bigl(\mathcal{H}^{\mathfrak{su}(2)_{-1}}_{\rm discr,+} \bigl)^{*} \cong \mathcal{H}^{\mathfrak{su}(2)_{-1}}_{\rm discr,-} \ .
\end{equation}

We point out that the $S$-matrix in (\ref{S_T_matrices}) is the same as the one for the affine $\widehat{\mathfrak{u}}(1)$ theory. Therefore, by applying the (continuous) Verlinde formula we obtain the following fusion rules for the finite-dimensional highest weight representations:
\begin{equation}\label{fusion}
    \mathcal{H}_{j} \times \mathcal{H}_{j'} \cong \mathcal{H}_{j+j'} ~~~~ \forall \, j,j' \in \tfrac{1}{2}\mathbb{N} \ .
\end{equation}
In particular, every module $\mathcal{H}_{j}$ is a simple current with infinite orbit. This will be relevant in Section \ref{sect_inv_from_coset}.

\subsection{Continuous representations}\label{sec_cont_repr}

We now consider the characters of the continuous representations (\ref{cont_su2__1_char}). Let us first compute the character of their spectrally flowed modules by (\ref{sigma}), which is
\begin{equation}\label{spec_flow_cont_char}
\begin{aligned}
       \sigma^{w}\bigl(\chi^{\lambda}_{j}\bigl)(t;\tau) &= q^{(j+\frac{1}{2})^{2}} \sum_{m \in \mathbb{Z}} x^{m+\lambda - \frac{w}{2}} \, q^{w(m+\lambda) -\frac{w^{2}}{4}} \, \frac{1}{\eta(\tau)^{3}} \\
    &= q^{(j+\frac{1}{2})^{2} + \frac{w^{2}}{4}} \sum_{m \in \mathbb{Z}} e^{2 \pi i m (\lambda - \frac{w}{2})} \, \delta(t + w \tau - m) \,\frac{1}{\eta(\tau)^{3}} \ .
\end{aligned}
\end{equation}
The modular $S$-transformation is then
\begin{equation}
\begin{aligned}
     \sigma^{w}&\bigl(\chi^{\lambda}_{j}\bigl)(\tfrac{t}{\tau},\tfrac{z}{\tau};-\tfrac{1}{\tau})  
    = e^{-\frac{2\pi i}{ \tau}\left((j+\frac{1}{2})^{2}+\frac{w^{2}}{4}\right)} \sum_{m \in \mathbb{Z} } e^{2 \pi i m (\lambda - \frac{w}{2})} \, \delta(\tfrac{t - w  -m \tau }{\tau}) \, \frac{(-i \tau)^{-\frac{3}{2}}}{\eta(\tau)^{3}} \\
    &= e^{\frac{-\pi i t^{2}}{2 \tau}} \,i\frac{|\tau|}{\tau} (-i \tau)^{-\frac{1}{2}}  \, e^{-\frac{2\pi i}{ \tau}(j+\frac{1}{2})^{2}} \sum_{m \in \mathbb{Z}} e^{2 \pi i (\lambda - \frac{w}{2})} \, q^{- \frac{m^{2}}{4}} \, \delta(t - w - m \tau) \, \frac{1}{\eta(\tau)^{3}} \\
    &= e^{-\frac{\pi i t^{2}}{2 \tau}} \sum_{w' \in \mathbb{Z}} \int_{0}^{1} d\lambda' \int_{\mathbb{R}} dj' \, S_{(j, \lambda, w),(j',\lambda',w')} \,  \sigma^{w'}\bigl(\chi^{\lambda'}_{j'}\bigl)(t ; \tau)  \ ,
\end{aligned}
\end{equation}
where in the second equality we used
\begin{equation}
    \delta \left(\frac{x}{\tau} \right) = |\tau|  \, \delta(x) ~~~~\forall \, x \in \mathbb{R} ~\forall \, \tau \in \mathbb{H} \ ,
\end{equation}
as well as (\ref{eta_modular_tr}). The modular $S$-matrix is given by
\begin{equation}\label{S_matrix}
    S_{(j,\lambda,w),(j',\lambda',w')} = \sqrt{2} \, e^{-4 \pi i(j+\frac{1}{2})(j'+\frac{1}{2})} \cdot i \, \frac{|\tau|}{\tau}  e^{-2 \pi i \left(w'\lambda + w \lambda'  + \frac{ww'}{2} \right)} \ ,
\end{equation}
which is symmetric and unitary:
\begin{equation}\label{S_matr_unitarity}
    \sum_{w' \in \mathbb{Z}} \int_{0}^{1} d \lambda' \, S_{(j,\lambda,w),(j',\lambda',w')} \, S_{(j',\lambda',w'),(j'',\lambda'',w'')}^{\dagger} = \delta(j-j'') \, \delta_{w,w''} \, \delta(\lambda - \lambda'') \ .
\end{equation}
The modular $T$-matrix transformation is easily computed to be  
\begin{equation}\label{T_tr}
    \sigma^{w}\bigl(\chi^{\lambda}_{j}\bigl)(t,z;\tau + 1) = e^{2 \pi i \left[(j+\frac{1}{2})^{2} + w(\lambda - \frac{w}{4}) - \frac{1}{8} \right]} \, \sigma^{w}\bigl(\chi_{j}^{\lambda}\bigl)(t,z;\tau) \ ,
\end{equation}
so that the $T$-matrix is also symmetric and unitary in the sense of (\ref{S_matr_unitarity}). 

From eq. (\ref{T_tr}) we see that $T$-invariance requires
\begin{equation}\label{T_inv_cond}
    (j+\tfrac{1}{2})^{2} + w(\lambda - \tfrac{w}{4}) = (j'+\tfrac{1}{2})^{2} + w'(\lambda' - \tfrac{w'}{4}) ~ \text{mod} \, 1 \ ,
\end{equation}
where the primed parameters correspond to the right-movers. Hence, by (\ref{S_matr_unitarity}) and (\ref{T_inv_cond}) it follows that the diagonal spectrum 
\begin{equation}\label{cont_diag_spectrum}
\mathcal{H}^{\mathfrak{su}(2)_{-1}}_{\rm diag} = \OplusInt_\mathbb{R} \, dj \, \bigoplus_{w \in \mathbb{Z}} \, \OplusInt_{\mathbb{R}/\mathbb{Z}} \, d \lambda \, \sigma^{w}\bigl(\mathcal{C}^{\lambda}_j\bigl) \otimes \, \overline{\sigma^{w}\bigl(\mathcal{C}^{\lambda}_j\bigl)} 
\end{equation}
is modular invariant. Note that here we used the fact that the set of $j$ and $\lambda$ for which $\lambda = j ~\text{mod}\,1$ and $j \in \tfrac{1}{2}\mathbb{Z}$, for which by (\ref{cont_char_special}) we have that $\sigma^{w}\bigl(\chi^{\lambda}_{j}\bigl)(t;\tau)$ is actually not equal to the character of $\sigma^{w} \bigl(\mathcal{C}^{\lambda}_j \bigl) $, has measure zero. 

By the symmetry 
\begin{equation}
    S_{(-j,-\lambda,-w),(-j',-\lambda',-w')} = S_{(j,\lambda,w),(j',\lambda',w')} \ ,
\end{equation}
also the charge conjugate spectrum
\begin{equation}\label{cont_cc_spectrum}
    \mathcal{H}^{\mathfrak{su}(2)_{-1}}_{\rm cc} = \OplusInt_\mathbb{R} \, dj \, \bigoplus_{w \in \mathbb{Z}} \, \OplusInt_{\mathbb{R}/\mathbb{Z}} \, d \lambda \, \sigma^{w}\bigl(\mathcal{C}^{\lambda}_j\bigl) \otimes \,\overline{\sigma^{-w}\bigl(\mathcal{C}^{-\lambda}_{-j}\bigl)} \ ,
\end{equation}
is modular invariant, where actually one can implement charge conjugation independently on $j$ and $(\lambda,w)$, so there are other two such invariants.

\subsection{Discrete partition functions}\label{sec_discr_inv}

Until now we found modular invariant spectra of $\mathfrak{su}(2)_{-1}$ containing continuous families of representations. However, when considering an $\mathfrak{su}(2)_{-1}$ WZW as coming from a sigma model on the compact Lie group ${\rm SU}(2)$, one wishes to impose a quantisation condition on the eigenvalues of $J^{3}_0$, therefore requiring the presence of only a discrete subset of representations.

If we only require $j \in \tfrac{1}{2}\mathbb{Z}$, we can use that (\ref{S_matrix}) readily factorises in a part consisting in the $\widehat{\mathfrak u}(1)$ $S$-matrix, and hence the following spectrum is modular invariant
\begin{equation}\label{su2__1_spec_mix}
    \mathcal{H}^{\mathfrak{su}(2)_{-1}}_{ j - {\rm cpt}} = \bigoplus_{\substack{j,\,j' \in \frac{1}{2} \mathbb{Z}: \\ j 
        = j' \, \text{mod}\,1}} \, \bigoplus_{w \in \mathbb{Z}} \, \OplusInt_{\mathbb{R}/\mathbb{Z}} d \lambda \, \sigma^{w}\bigl(\mathcal{C}^{\lambda}_j\bigl) \otimes \, \overline{\sigma^{w}\bigl(\mathcal{C}^{\lambda}_{j'}\bigl)} \ ,
\end{equation}
as well as the $(\lambda,w)$ charge conjugate version of it.

We now look for fully discrete invariants obtained by imposing both quantisation conditions
\begin{equation}\label{lambda_j_quantisation}
    {\lambda = 0, \tfrac{1}{2}} \qquad \text{and} \qquad j \in \tfrac{1}{2}\mathbb{Z} \ .
\end{equation}
Then eq. (\ref{T_inv_cond}) leads to the following combinations (where we omit an implicit summation over $j , j' \in \tfrac{1}{2}\mathbb{Z}$ and $w, w' \in \mathbb{Z}$):
\begin{equation}\label{part_functs}
    \begin{aligned}
        Z^{\mathfrak{su}(2)_{-1}}_1(t;\tau) & = \sum_{j - j' = \frac{w- w'}{2} \, \text{mod}\, 1\,}  (-1)^{2(j-j')} \, \sigma^{w}\bigl(\chi_{j}^j\bigl)(t;\tau) \, \overline{\sigma^{w'}\bigl(\chi_{j'}^{ j'}\bigl)(t;\tau)} \ ,\\
        Z^{\mathfrak{su}(2)_{-1}}_2(t;\tau) & = \sum_{\substack{j = j' \, \text{mod}\, 1, \\ w=w' \, \text{mod} \, 2}} \sum_{\lambda = 0, \frac{1}{2}} \, \sigma^{w}\bigl(\chi_{j}^{ \lambda}\bigl)(t;\tau) \, \overline{\sigma^{w'}\bigl(\chi_{j'}^{ \lambda}\bigl)(t;\tau)}  
            \ ,
    \end{aligned}
\end{equation}
which one checks to be indeed modular invariant. It is important to point out that for (\ref{lambda_j_quantisation}) the character function $\sigma^{w}\bigl(\chi_{j}^{ \lambda}\bigl)$ differ from the character of $\sigma^{w} \bigl ( \mathcal{C}_{j}^{ \lambda} \bigl)$. Hence, a decomposition of (\ref{part_functs}) in terms of $\mathfrak{su}(2)_{-1}$ affine characters is not so direct. Also, note that the alternating sign in $Z^{\mathfrak{su}(2)_{-1}}_1(t;\tau)$ makes it difficult to interpret it as a partition function for $\mathfrak{su}(2)_{-1}$. In the next Section we address these issues and give an interpretation of these modular invariants in relation to the $\mathfrak{su}(2)_{-1}$ WZW model.

\section{Understanding the discrete invariants}
\label{sect_inv_from_coset}

In this section we analyze the discrete invariants of eq. (\ref{part_functs}) and give two different but related interpretations. The first one is in terms of a coset of $\mathfrak{psu}(2|2)_1$ (or equivalently of $\mathfrak{u}(2|2)_1$); this nicely relates to some observations made in \cite{Gaberdiel:2023nhb}. The second one comes from vertex operator (super)algebra extensions of $\mathfrak{su}(2)_{-1}$. In the latter picture,
we find that the first invariant in (\ref{part_functs}) corresponds to a supersymmetric rational CFT whose characters possess the same representation of the modular group as $\mathfrak{su}(2)_{1}$. The second invariant in  (\ref{part_functs}) corresponds instead to a rational (bosonic) CFT containing eight admissible representations.

Recall that there is a conformal embedding \cite{Conf_emb,Goddard:1987td}
\begin{equation}\label{psu22_conf_emb}
    \mathfrak{su}(2)_{-1} \oplus \mathfrak{su}(2)_{1} \subset \mathfrak{psu}(2|2)_{1} \ .
\end{equation} 
From the point of view of $\mathfrak{su}(2)_{-1}$, it is therefore fairly natural to consider the coset construction $\mathfrak{psu}(2|2)_1/\mathfrak{su}(2)_1$. Moreover, discrete modular invariants of $\mathfrak{psu}(2|2)_{1}$ are known \cite[(5.11),(5.17)]{Gaberdiel:2023nhb}:
\begin{equation}\label{psu22_spectra}
\begin{aligned}
    \mathcal{H}^{\mathfrak{psu}(2|2)_{1}}_1 &= \bigoplus_{w, w' \in \mathbb{Z}} \tilde{\sigma}^{w}\bigl(\mathscr{L}\bigl) \otimes \, \overline{\tilde{\sigma}^{w'}\bigl(\mathscr{L}\bigl)} \ , \\
    \mathcal{H}^{\mathfrak{psu}(2|2)_{1}}_2 &= \bigoplus_{\substack{w, w' \in \mathbb{Z} :\\ w= w' \, \text{mod} \, 2}} \bigoplus_{\lambda = 0, \frac{1}{2}} \tilde{\sigma}^{w}\bigl(\mathscr{F}_{\lambda}\bigl) \otimes \, \overline{\tilde{\sigma}^{w'}\bigl(\mathscr{F}_{\lambda}\bigl)}  \ ,
\end{aligned}
\end{equation}
where $\tilde{\sigma}$ is the spectral flow of $\mathfrak{psu}(2|2)_1$ and according to our notation 
\begin{equation}\label{sigma_tilde}
    \tilde{\sigma} = \sigma \circ \sigma_K
\end{equation} 
with $\sigma$ as in (\ref{sigma}) and $\sigma_K$ denoting the spectral flow of $\mathfrak{su}(2)_1$.

We show that the invariants (\ref{part_functs}) can be obtained by inserting the branching rules of the embedding (\ref{psu22_conf_emb}) and then passing to the coset CFT by quotienting out $\mathfrak{su}(2)_{1}$. Recall that the modular invariant of $\mathfrak{su}(2)_{1}$ is diagonal and it contains the only two integrable representations of spin $\ell=0,1/2$ whose characters we denote by $\chi^{(1)}_\ell(z;\tau)$. Hence, we conjecture that a general mass matrix for the $\mathfrak{su}(2)_{-1}$ theory is given by 
\begin{equation}\label{coset_mass_matr}
    \mathcal{M}^{\mathfrak{su}(2)_{-1}}_{(j,w), \, (j',w')} = \sum_{\ell=0,  \frac{1}{2}} \mathcal{M}^{\mathfrak{su}(2)_{-1} \oplus \, \mathfrak{su}(2)_{1}}_{(j,w,\ell), \, (j',w',\ell)} \ ,
\end{equation}
where by (\ref{psu22_conf_emb}) the mass matrix on the right-hand side is essentially that of $\mathfrak{psu}(2|2)_1$.

\subsection{The first invariant}

Let us start with the $\mathfrak{psu}(2|2)_{1}$ modular invariant $\mathcal{H}^{\mathfrak{psu}(2|2)_1}_1$ in (\ref{psu22_spectra}), where\footnote{Note that when considering modular invariance on WZW models superalgebras one considers supercharacters rather then characters, see \cite{Gaberdiel:2023nhb}. In particular, the modular invariant partition function of $\mathcal{H}^{\mathfrak{psu}(2|2)_1}_1$ is
\begin{equation}
    \sum_{w,w'\in \mathbb{Z}} \text{sch}\bigl[\tilde{\sigma}^{w}\bigl(\mathscr{L}\bigl)](t,z;\tau) \, \overline{\text{sch}\bigl[\tilde{\sigma}^{w'}\bigl(\mathscr{L}\bigl)](t,z;\tau)} \ .
\end{equation}}
\begin{equation}\label{psu22_L_char}
\begin{aligned}
        \text{sch}\bigl[\mathscr{L} \bigl](t,z;\tau) 
    &= \chi^{(1)}_{0}(z;\tau) \sum_{j \in \mathbb{N}} (2 j +1) \, \text{ch}\bigl[\mathcal{H}_j\bigl] (t;\tau) - \chi^{(1)}_{\frac{1}{2}}(z;\tau) \sum_{j \in \mathbb{N} + \frac{1}{2}} (2 j +1) \, \text{ch}\bigl[\mathcal{H}_j\bigl](t;\tau) \\
    &= \chi^{(1)}_{0}(z;\tau) \, \text{ch}\bigl[\mathbb{H}_0\bigl](t;\tau) - \chi^{(1)}_{\frac{1}{2}}(z;\tau) \, \text{ch}\bigl[\mathbb{H}_{\frac{1}{2}}\bigl](t;\tau) \ ,
\end{aligned}
\end{equation}
where we introduced the \textit{extended} $\mathfrak{su}(2)_{-1}$ characters
\begin{equation}\label{su2_1_extended_characters}
\begin{aligned}
        \text{ch}\bigl[\mathbb{H}_0\bigl](t;\tau) &= \sum_{j \in \mathbb{N}}(2j+1) \, \text{ch}\bigl[\mathcal{H}_j\bigl](t;\tau) = \frac{\partial_{t} \vartheta_{2}(t;2\tau)}{\pi \vartheta_{1}(t;\tau)} \ , \\
        \text{ch}\bigl[\mathbb{H}_{\frac{1}{2}}\bigl](t;\tau) &= \sum_{j \in \mathbb{N}+\frac{1}{2}}(2j+1) \, \text{ch}\bigl[\mathcal{H}_j\bigl](t;\tau)  = \frac{\partial_{t} \vartheta_{3}(t;2\tau)}{\pi \vartheta_{1}(t;\tau)} \ ,
\end{aligned}
\end{equation}
and the meromorphic expressions hold on the same convergence region as for (\ref{character_j_finite_dim}). 
Analogously, for the spectrally flowed modules we compute 
\begin{equation}\label{sigma_L_char}
\begin{aligned}
    \text{sch}\bigl[\tilde{\sigma}^{w}(\mathscr{L}) \bigl](t,z;\tau) &=  \chi^{(1)}_{[\frac{w}{2}]}(z;\tau) \, \text{ch}\bigl[\sigma^w \bigl(\mathbb{H}_{0} \bigl)\bigl](t;\tau)  -  \chi^{(1)}_{[\frac{w+1}{2}]}(z;\tau) \, \text{ch}\bigl[\sigma^w \bigl(\mathbb{H}_{\frac{1}{2}} \bigl)\bigl](t;\tau)  \ ,
\end{aligned}
\end{equation}
where we used that $\sigma_K$ is an involution mapping $\chi^{(1)}_0$ and $\chi^{(1)}_{1/2}$ onto each other, and introduced the notation $[a] = a~\text{mod}\,1$ for $a \in \mathbb{Q}$. Hence, (\ref{coset_mass_matr}) yields
\begin{equation}\label{su2_1_spectrum}
\begin{aligned}
        \mathcal{H}^{\mathfrak{su}(2)_{-1}}_1 &= \bigoplus_{\substack{w, w' \, \in \, \mathbb{Z}  \, : \\ w = w' \, \text{mod} \, 2}} \bigoplus_{\ell =0,  \frac{1}{2}} \sigma^{w}\bigl(\mathbb{H}_{\ell}\bigl) \otimes \, \overline{\sigma^{w'}\bigl(\mathbb{H}_{\ell}\bigl)} \oplus \left( \bigoplus_{\substack{w, w' \, \in \, \mathbb{Z} \, : \\ w = w' + 1 \, \text{mod} \, 2}} \bigoplus_{\ell=0, \frac{1}{2}} \sigma^{w}\bigl(\mathbb{H}_{\ell}\bigl)  \otimes \, \overline{\sigma^{w'}\bigl(\mathbb{H}_{\ell+\frac{1}{2}}\bigl)} \right)  \\
        &= \mathbb{L} \otimes \overline{\mathbb{L}} \oplus \sigma\bigl(\mathbb{L}\bigl) \otimes \, \overline{\sigma\bigl(\mathbb{L}\bigl)} \ ,
\end{aligned}
\end{equation}
where we defined the \textit{extended} $\mathfrak{su}(2)_{-1}$ module
\begin{equation}\label{X_module}
\begin{aligned}
    \mathbb{L} &= \bigoplus_{w \, \text{even}} \sigma^{w}\bigl(\mathbb{H}_{0}\bigl) \oplus \bigoplus_{w \, \text{odd}} \sigma^{w}\bigl(\mathbb{H}_{\frac{1}{2}}\bigl) \\
    & \cong \bigoplus_{j \in \mathbb{N}} (2j +1) \bigoplus_{w \, \text{even}} \sigma^{w}\bigl(\mathcal{H}_{j}\bigl) \oplus \bigoplus_{j \in \mathbb{N} + \frac{1}{2}} (2j +1) \bigoplus_{w \, \text{odd}} \sigma^{w}\bigl(\mathcal{H}_{j}\bigl) \ ,
\end{aligned}
\end{equation}
and the supercharacters read
\begin{equation}\label{X_characters}
\begin{aligned}
       \text{sch}\bigl[\mathbb{L}\bigl] &= \sum_{w \, \text{even}} \text{ch}\bigl[\sigma^w \bigl(\mathbb{H}_{0} \bigl)\bigl] - \sum_{w \, \text{odd}} \text{ch}\bigl[\sigma^w \bigl(\mathbb{H}_{\frac{1}{2}} \bigl)\bigl] \ ,\\ \text{sch}\bigl[\sigma\bigl(\mathbb{L}\bigl)\bigl] &= \sum_{w \, \text{odd}} \text{ch}\bigl[\sigma^w \bigl(\mathbb{H}_{0} \bigl)\bigl] - \sum_{w \, \text{even}} \text{ch}\bigl[\sigma^w \bigl(\mathbb{H}_{\frac{1}{2}} \bigl)\bigl] \ . 
\end{aligned}
\end{equation}
The last ingredient for connecting the invariant $Z_1^{\mathfrak{su}(2)_{-1}}$ to $\mathcal{H}_1^{\mathfrak{psu}(2|2)_1}$ is obtained by
combining (\ref{cont_char_special}) and (\ref{cont_mod_structure}), which yields
\begin{equation}\label{inf_sum__chi_j}
\begin{aligned}
        \sum_{j \in \mathbb{Z}}  \chi_{j}^{j} &= \sum_{j \in \mathbb{N} + \frac{1}{2} } (2j+1) \, \text{ch}\bigl[\mathcal{C}_{j-\frac{1}{2}}^{j-\frac{1}{2}}\bigl] = \text{ch}\bigl[\sigma\bigl(\mathbb{H}_{\frac{1}{2}} \bigl)\bigl]  + \, \text{ch}\bigl[\sigma^{-1}\bigl(\mathbb{H}_{\frac{1}{2}} \bigl)\bigl] + \, 2 \, \text{ch}\bigl[\mathbb{H}_{0}\bigl] \ , \\
        \sum_{j \in \mathbb{Z} + \frac{1}{2}}  \chi_{j}^{j} &= \sum_{j \in \mathbb{N} } (2j+1) \, \text{ch}\bigl[\mathcal{C}_{j-\frac{1}{2}}^{j-\frac{1}{2}}\bigl] = \text{ch}\bigl[\sigma\bigl(\mathbb{H}_{0} \bigl)\bigl]  + \, \text{ch}\bigl[\sigma^{-1}\bigl(\mathbb{H}_{0} \bigl)\bigl] + \, 2 \, \text{ch}\bigl[\mathbb{H}_{\frac{1}{2}}\bigl] \ , 
\end{aligned}
\end{equation}
as equations of formal power series. At this point, one easily verifies that the partition function $Z^{\mathfrak{su}(2)_{-1}}_1(t;\tau)$ of (\ref{part_functs}) corresponds to (\ref{su2_1_spectrum}). 

As previously mentioned, the presence of minus signs in the partition function $Z^{\mathfrak{su}(2)_{-1}}_1$ as well as in (\ref{X_characters}), prevents us from interpreting this invariant directly in terms of a $\mathfrak{su}(2)_{-1}$ WZW model. Nevertheless, there is a natural framework that enables us to bypass this issue: that of extended algebras. In fact, (\ref{X_characters}) may be interpreted as supercharacters of irreducible modules of an extended algebra of $\mathfrak{su}(2)_{-1}$, which we expect to be a vertex operator superalgebra. In favor of this, we identify $\mathbb{L}$ with the vacuum module of the extension of $\mathfrak{su}(2)_{-1}$ by the fields corresponding to 
\begin{equation}\label{extended_modules}
    \mathcal{H}_{1} \ ,~ \sigma^{2}\bigl(\mathcal{H}_{0}\bigl) \ ,~ \sigma^{-2}\bigl(\mathcal{H}_{0}\bigl) ~~~~\text{and}~~~~\sigma\bigl(\mathcal{H}_{\frac{1}{2}}\bigl) \ ,~\sigma^{-1}\bigl(\mathcal{H}_{\frac{1}{2}}\bigl) \ ,
\end{equation}
where we expect the corresponding fields to be bosonic in the first three cases and fermionic in the second two. All of these modules have states with integer conformal dimensions, which is more generally true for every $\sigma^{w}\bigl(\mathcal{H}_{j}\bigl)$ with $j \in \mathbb{N}$ and $w$ even or with $j \in \mathbb{N} + \tfrac{1}{2}$ and $w$ odd. Together with (\ref{fusion}), we expect that the modules in (\ref{extended_modules}) are simple currents. Then, on the extended algebra $\sigma$ becomes an involution, i.e. $\sigma^{2}\bigl(\mathbb{L}\bigl) \cong \mathbb{L}$, so that the irreducible modules of this theory are $\mathbb{L}$ and $\sigma\bigl(\mathbb{L}\bigl)$ and (\ref{su2_1_spectrum}) is the corresponding diagonal invariant. Moreover, by (\ref{sigma_L_char}) we have that
\begin{equation}\label{extended_psu22_module}
    \sum_{w \in \mathbb{Z}} \text{sch}\bigl[\sigma^{w}\bigl(\mathscr{L}\bigl) \bigl](z,t;\tau) = \chi^{(1)}_0(z;\tau) \, \text{sch}\bigl[\mathbb{L}\bigl](t;\tau) + \chi^{(1)}_{\frac{1}{2}} (z;\tau) \,\text{sch}\bigl[\sigma\bigl(\mathbb{L}\bigl)\bigl](t;\tau) \ ,
\end{equation}
which by modular invariance of $\mathcal{H}^{\mathfrak{psu}(2|2)_1}_1$ must be $S$-invariant by itself. This in turn forces the S-matrix $\mathbb{S}^{(1)}$ of the extended theory in the ordered basis $(\mathbb{L}, \sigma\bigl(\mathbb{L}\bigl))$ to be the same of that of $\mathfrak{su}(2)_1$, namely
\begin{equation}\label{su2_1_S_matr}
    \mathbb{S}^{(1)} = \frac{1}{\sqrt{2}} \begin{pmatrix}
        1 & 1 \\ 1 & -1
    \end{pmatrix} \ .
\end{equation}
This is consistent with the interpretation given in \cite[(5.18)]{Gaberdiel:2023nhb}, where (\ref{extended_psu22_module}) is seen as the vacuum module of a simple current extension of $\mathfrak{psu}(2|2)_1$. Here we have quotiented out the $\mathfrak{su}(2)_1$, which then yields an invariant for a simple current extension of $\mathfrak{su}(2)_{-1}$.

We point out that one can also start with the $\mathfrak{psu}(2|2)_1$ invariant given in \cite[(5.10)]{Gaberdiel:2023nhb}. By comparing (\ref{psu22_L_char}) with (\ref{psu22_F_12_char}), it immediately follows that also the $\mathfrak{su}(2)_{-1}$ spectrum obtained from (\ref{su2_1_spectrum}) by replacing $\mathcal{H}_j$ with $\mathcal{C}_j^j$ in (\ref{X_module}) is (super)modular invariant. In fact, the corresponding extended modules are the irreducibles of the same $\mathfrak{su}(2)_{-1}$ extension by (\ref{extended_modules}), obtained from the orbit of $\mathcal{C}^0_0$ rather then that of $\mathcal{H}_0$. Then, the extended modules have the same structure as (\ref{X_module}) by replacing $\mathcal{H}_j$ with~$\mathcal{C}^j_j$.

\subsection{The second invariant}

We now look at the same construction (\ref{coset_mass_matr}) but starting from the $\mathfrak{psu}(2|2)_{1}$ invariant $\mathcal{H}^{\mathfrak{psu}(2|2)_1}_2$ of (\ref{psu22_spectra}). The relevant supercharacters are \cite{Eberhardt:2018ouy}
\begin{equation}\label{F_lambda_char}
\begin{aligned}
\text{sch}\bigl[\mathscr{F}_{\lambda}\bigl](t,z;\tau) 
    &= \chi^{(1)}_{0}(z;\tau) \sum_{j \in \mathbb{Z}}  \chi_{j}^{j+\lambda}(t;\tau) - \chi^{(1)}_{\frac{1}{2}}(z;\tau) \sum_{j \in \mathbb{Z} + \frac{1}{2}} \chi_{j}^{j+\lambda}(t;\tau) \ 
\end{aligned}
\end{equation}
$\text{for}~\lambda \in [0,1)$. In particular, we can write
\begin{equation}\label{psu22_F_12_char}
\begin{aligned}
\text{sch}\bigl[\mathscr{F}_{0}\bigl](t,z;\tau) 
    =& \ \chi^{(1)}_{0}(z;\tau) \sum_{j \in \mathbb{N} + \frac{1}{2}} (2 j +1) \, \text{ch}\bigl[\mathcal{C}_{j-\frac{1}{2}}^{j-\frac{1}{2}}\bigl](t;\tau)  \\
    & - \, \chi^{(1)}_{\frac{1}{2}}(z;\tau) \sum_{j \in \mathbb{N} } (2 j +1) \, \text{ch}\bigl[\mathcal{C}_{j-\frac{1}{2}}^{j-\frac{1}{2}}\bigl](t;\tau) \ ,
\end{aligned}
\end{equation}
where we used eq. (\ref{cont_char_special}). 
Then, the $\mathfrak{su}(2)_{-1}$ invariant obtained from $\mathcal{H}^{\mathfrak{psu}(2|2)_1}_2$ is then
\begin{equation}\label{su2_1_spectrum2}
    \mathcal{H}^{\mathfrak{su}(2)_{-1}}_2 =  \bigoplus_{\ell=0,\frac{1}{2}} \bigoplus_{\lambda = 0 , \frac{1}{2}}  \mathbb{C}_{\ell}^{\lambda} \otimes \overline{\mathbb{C}_{\ell}^{\lambda}} \oplus  \sigma \bigl(\mathbb{C}_{\ell}^{\lambda}\bigl) \otimes \, \overline{\sigma \bigl(\mathbb{C}_{\ell}^{\lambda}\bigl)} \ ,
\end{equation}
where we defined the \textit{extended} modules
\begin{equation}\label{C0_modules}
    \mathbb{C}^{0}_{\ell} = \bigoplus_{j \in \mathbb{N} + \frac{1}{2} - \ell} (2j+1) \bigoplus_{w \, \text{even}} \sigma^{w}\bigl(\mathcal{C}^{j-\frac{1}{2}}_{j-\frac{1}{2}}\bigl) ~~~~ \text{for} \ \ell = 0,\tfrac{1}{2} \ ,
\end{equation}
and
\begin{equation}\label{C1_modules}
\mathbb{C}^{\frac{1}{2}}_{\ell} = \begin{cases}
        \bigoplus\limits_{j \in \mathbb{Z} } \bigoplus\limits_{w \, \text{even}} \sigma^{w}\bigl( \mathcal{C}^{j+\frac{1}{2}}_{j}\bigl) = 2 \bigoplus\limits_{j \in \mathbb{N} } \bigoplus\limits_{w \, \text{even}} \sigma^{w}\bigl( \mathcal{C}^{j+\frac{1}{2}}_{j}\bigl) ~~~~~~~~~~~~~~~~~~~~~~~~~~~~~~~\, \ell = 0 \ , \\
        \bigoplus\limits_{j \in \mathbb{Z}+\frac{1}{2} } \bigoplus\limits_{w \, \text{even}} \sigma^{w}\bigl( \mathcal{C}^{j+\frac{1}{2}}_{j}\bigl) = \bigoplus\limits_{w \, \text{even}} \sigma^{w}\bigl(\mathcal{C}^{0}_{-\frac{1}{2}}\bigl) \oplus \, 2 \bigoplus\limits_{j \in \mathbb{N} + \frac{1}{2} } \bigoplus\limits_{w \, \text{even}} \sigma^{w}\bigl(\mathcal{C}^{j+\frac{1}{2}}_{j}\bigl) ~~~~ \ell = \tfrac{1}{2} \ .
    \end{cases} 
\end{equation}
Using (\ref{inf_sum__chi_j}) it is straightforward to check that $Z^{\mathfrak{su}(2)_{-1}}_2(t;\tau)$ from (\ref{part_functs}) is the partition function corresponding to (\ref{su2_1_spectrum2}). In contrast to (\ref{su2_1_spectrum}), in (\ref{su2_1_spectrum2}) there is no hint of supersymmetry at the level of characters. Indeed, we can interpret the modules $\mathbb{C}^{\lambda}_{\ell}$ as modules of the extension of $\mathfrak{su}(2)_{-1}$ by the (bosonic) simple currents
\begin{equation}\label{extension2}
    \mathcal{H}_1 \ , ~ \sigma^{2} \bigl(\mathcal{H}_0 \bigl) \ , ~ \sigma^{-2} \bigl(\mathcal{H}_0 \bigl) \ .
\end{equation}
In particular, for $\lambda, \ell = 0, \tfrac{1}{2}$ the module $\mathbb{C}^{\lambda}_{\ell}$ is the fusion orbit of $\mathcal{C}^{\ell + \lambda}_{\ell}$ under the currents (\ref{extension2}). As above, $\sigma$ becomes an involution on the extended modules, i.e. $\sigma^2\bigl(\mathbb{C}^{\lambda}_{\ell} \bigl) \cong \mathbb{C}^{\lambda}_{\ell}$. Therefore, this extended theory is a rational CFT consisting of eight admissible representations and (\ref{su2_1_spectrum2}) is the corresponding diagonal spectrum. One can compute the $S$-matrix of this extended theory, which is
\begin{equation}\label{S_matrix_interesting}
    \mathbb{S}^{(2)}_{(\ell,\lambda,r),(\ell',\lambda',r')} = \frac{i}{\sqrt{2}} \frac{|\tau|}{\tau} \, (-1)^{(2\ell + 1 + r)(2 \ell' + 1 + r') + 2\lambda r' + 2 \lambda' r} \ ,
\end{equation}
where $(\ell, \lambda , r)$ labels the extended module $\sigma^{r}\bigl(\mathbb{C}^{\lambda}_{\ell} \bigl)$ with $\lambda,\ell = 0, \tfrac{1}{2}$ and $r = 0,1$.

Lastly, we mention that one can also consider the continuous diagonal (and charge conjugate) $\mathfrak{psu}(2|2)_1$ invariants in \cite[(5.4-5)]{Gaberdiel:2023nhb}. Then, by (\ref{F_lambda_char}) the coset construction yields exactly the modular invariant $\mathfrak{su}(2)_{-1}$ spectrum (\ref{su2__1_spec_mix}) (and its charge conjugate version). In conclusion, from the coset construction of $\mathfrak{psu}(2|2)_1$ we retrieved all the $\mathfrak{su}(2)_{-1}$ invariants found in Section \ref{sec_cont_repr} and \ref{sec_discr_inv} with quantised half-integer spins.

\section{A free field invariant}
\label{section_free_field_invariant}

In this section we propose a modular invariant for the free field theory of four symplectic bosons, answering another question raised \cite{Gaberdiel_2018}. The subtle point in such construction is related to the remark at the beginning of Section \ref{section_characters}: the association of modules to character functions is not injective. More specifically, infinite orbits under the spectral flows $\sigma$ (\ref{sigma}) and $\sigma_U$ (\ref{sigma_U__}) possess the same character function, see (\ref{section_free_field_characters}). Moreover, as for the partition function of free fermions, where one has to sum over the different spin structures, also for the case of the symplectic bosons one has to consider the different possible boundary conditions of the free fields on the torus. In particular, we have to require that both pairs of symplectic bosons possess the same boundary condition. At the level of characters, considering different boundary condition has the effect of introducing the operator $(-1)^B$, where $B$ is the boson number. More explicitely, from (\ref{NS_level_1}) we compute
\begin{equation}\label{char_bos_NS}
     \text{ch}\bigl[ \mathcal{V}  \bigl](t,\mu;\tau) = \frac{\eta(\tau)^{2}}{\vartheta_{4}(\frac{t+\mu}{2};\tau)\vartheta_{4}(\frac{t-\mu}{2};\tau)} \longrightarrow \frac{\eta(\tau)^{2}}{\vartheta_{3}(\frac{t+\mu}{2};\tau)\vartheta_{3}(\frac{t-\mu}{2};\tau)} \ ,
\end{equation}
valid for $|q|<|x|,|y|<|q|^{-1}$, and from (\ref{R_character})
\begin{equation}\label{char_bos_R}
     \text{ch}\bigl[ \mathcal{R}^{+}  \bigl](t,\mu;\tau) = \frac{\eta(\tau)^{2}}{\vartheta_{1}(\frac{t+\mu}{2};\tau)\vartheta_{1}(\frac{t-\mu}{2};\tau)} \longrightarrow \frac{\eta(\tau)^{2}}{\vartheta_{2}(\frac{t+\mu}{2};\tau)\vartheta_{2}(\frac{t-\mu}{2};\tau)} \ ,
\end{equation}
valid for $|q|<|x|<|q|^{-1}$ and $|q|^{2} <|y|<1$. See Appendix \ref{theta_functions} for our conventions on theta functions. In particular,
\begin{equation}\label{tilde_char}
    \text{ch}\bigl[ \mathcal{V}_U  \bigl](t;\tau) \longrightarrow  (-1)^{2U} \text{ch}\bigl[ \mathcal{V}_U  \bigl](t;\tau) \ , \  \text{ch}\bigl[ \mathcal{R}^{+}_U  \bigl](t;\tau) \longrightarrow (-1)^{2U} \text{ch}\bigl[ \mathcal{R}^{+}_U  \bigl](t;\tau) \ ,
\end{equation}
where recall that the subscript denotes the subrepresentation at fixed $U_0=U \in \tfrac{1}{2}\mathbb{Z}$.

We now look at the action of the spectral flow on the free field modules. In contrast to the case of free fermions, where the spectral flows are involutions, for symplectic bosons spectral flows have infinite order and map free field character functions expanded on (generally) disjoint convergence regions one onto the other. Their action on characters is given by (\ref{sigma_action_modules}) and the analogous equation for $\sigma_{U}$ with $y$, $\mu$ at the place of~$x$,~$t$. 
One can for example compute from (\ref{NS_dec_char}) 
\begin{equation}\label{sigma_U_char_U}
    \text{ch}\bigl[ \sigma^{w}_U \bigl(\mathcal{V}_U \bigl) \bigl](t,\mu;\tau) = \frac{q^{-U^{2}}}{\eta(\tau)} \sum_{j \in \mathbb{N} + |U - \frac{w}{2}|} \text{ch}\bigl[\mathcal{H}_j  \bigl](t;\tau) \ ,
\end{equation}
and
\begin{equation}\label{sigma_char_U}
     \text{ch}\bigl[ \sigma^{w} \bigl(\mathcal{V}_U \bigl) \bigl](t,\mu;\tau) = \frac{q^{-U^{2}}}{\eta(\tau)} \sum_{j \in \mathbb{N} + |U|} \text{ch}\bigl[ \sigma^{w} \bigl(\mathcal{H}_j \bigl)  \bigl](t;\tau) \ .
\end{equation}
For what concerns the characters of the full free field representations, with (\ref{theta_function_periodicity2}) we compute
\begin{equation}\label{chi_NS_U}
    \text{ch}\bigl[ \sigma_{U}^{w}\bigl( \mathcal{V} \bigl)  \bigl](t,\mu;\tau) = \frac{\eta(\tau)^{2}}{\vartheta_{i}(\frac{t+\mu}{2};\tau)\vartheta_{i}(\frac{t-\mu}{2};\tau)} ~~~~ \begin{cases}
        i=1~~\text{if}~w~\text{even} \\ i=4~~\text{if}~w~\text{odd} \ ,
        \end{cases}
\end{equation}
with convergence region $|q|<|x|<|q|^{-1}$ and $|q|^{1-w}<|y|<|q|^{-1-w}$. 
Similarly,
\begin{equation}\label{chi_NS_sigma}
    \text{ch}\bigl[ \sigma^{w}\bigl( \mathcal{V} \bigl)  \bigl](t,\mu;\tau) = (-1)^{w}\frac{\eta(\tau)^{2}}{\vartheta_{i}(\frac{t+\mu}{2};\tau)\vartheta_{i}(\frac{t-\mu}{2};\tau)} ~~~~ \begin{cases}
        i=1~~\text{if}~w~\text{even} \\ i=4~~\text{if}~w~\text{odd} \ ,
        \end{cases}
\end{equation}
valid for $|q|^{1-w}<|x|<|q|^{-1-w}$ and $|q|<|y|<|q|^{-1}$. Similarly for the other boundary condition on the torus, following the transformations of (\ref{char_bos_R}) and (\ref{tilde_char}), which do not affect the convergence region. Note that these characters (generally) possess disjoint convergence regions in the $(x,y)$-plane, hence character functions clearly do not detect the spectral flow action.

Ignoring momentarily the convergence issues of characters and the spectral flow action, we consider simply character functions. Then we find the modular invariant
\begin{equation}\label{sp4_inv}
        Z^{\mathfrak{sp}(4)_{-1/2}}(t,\mu;\tau) = \frac{1}{2} \sum_{i=1}^{4} \left| \frac{\eta(\tau)^{2}}{\vartheta_{i}(\frac{t+\mu}{2};\tau)\vartheta_{i}(\frac{t-\mu}{2};\tau)} \right|^{2} \ ,
\end{equation}
which can be interpreted as an invariant for the $\mathfrak{sp}(4)_{-1/2}$ theory. Note in fact that (\ref{sp4_inv}) is analogous to the $\mathfrak{so}(4)_1$ invariant obtained from the free field realisation in terms of four fermions. Even though (\ref{sp4_inv}) surely is a modular invariant function, we suspect that in order to associate to it an actual module content we must include spectrally flowed representations. As we mentioned, the character functions appearing in (\ref{sp4_inv}) are blind to the action of $\sigma$ and $\sigma_U$, see eqs. (\ref{chi_NS_U}) and (\ref{chi_NS_sigma}).

Our approach for finding a free field invariant is to start from \cite[(5.21)]{Gaberdiel:2023nhb}, which gives a free field invariant for the theory of four symplectic bosons and four fermions. The four fermions yield $\mathfrak{so}(4)_1 \cong \mathfrak{su}(2)_{1}\oplus \mathfrak{su}(2)_{1}$, and we expect that quotienting out their contribution we obtain an invariant for only the symplectic bosons. Guided by this philosophy, we propose the spectrum
\begin{equation}\label{proposal}
    \bigoplus_{\substack{w,s,\bar{w},\bar{s}\in \mathbb{Z}:\\ w = \bar{w} \, \text{mod}\,2, \\ s = \bar{s} \, \text{mod}\,2}} \sigma^{w} \sigma_U^{s} \bigl(\mathcal{R}^{+} \bigl) \otimes \, \overline{\sigma^{\bar w} \sigma_U^{\bar s} \bigl(\mathcal{R}^{+} \bigl)} \ ,
\end{equation}
where recall that $\mathcal{R}^{+}$ is the R sector defined by (\ref{sympl_bos_zero}) with $m_1,m_2 \in \tfrac{1}{2}\mathbb{N}_0$.
The full partition function is obtained from (\ref{proposal}) by summing over the two possible boundary conditions for the free boson fields on the torus, one contributes to the terms with $i=1,4$ in (\ref{sp4_inv}) and the other to those with $i=2,3$.

Let us now prove more explicitly modular invariance of the proposed free field spectrum.
We start by computing
\begin{equation}\label{speculation}
\begin{aligned}
        \text{ch}&\bigl[\bigoplus_{\substack{r,w\in \mathbb{Z}: \\ r = w \, \text{mod}\,2 }}  \sigma^{w}\sigma^{r}_U \bigl(\mathcal{R}^+ \bigl)\bigl]  = \sum_{\substack{r,w \in \mathbb{Z}:\\ r = w \, \text{mod}\,2}} (-1)^w \sum_{U \in \frac{1}{2}\mathbb{Z}} \frac{q^{-U^2}}{\eta(\tau)} \sum_{j \in \mathbb{N} + |U+\frac{r-1}{2}|} \text{ch}\bigl[\sigma^{w} \bigl(\mathcal{H}_j \bigl)\bigl] \\
        & = \sum_{U \in \frac{1}{2}\mathbb{Z}} \frac{q^{-U^2}}{\eta(\tau)} \left(\sum_{j \in \mathbb{N} + [U + \frac{1}{2}]} \sum_{w \, \text{even}} (2j+1) \, \text{ch}\bigl[\sigma^{w} \bigl(\mathcal{H}_j \bigl)\bigl] \, - \sum_{j \in \mathbb{N} + [U ]} \sum_{w \, \text{odd}} (2j+1) \, \text{ch}\bigl[\sigma^{w} \bigl(\mathcal{H}_j \bigl)\bigl]  \right) \\
        & = \sum_{U \in \frac{1}{2}\mathbb{Z}} \frac{q^{-U^2}}{\eta(\tau)} (-1)^{2U+1} \, \text{sch}\bigl[\sigma^{2[U+\frac{1}{2}]}\bigl(\mathbb{L} \bigl) \bigl] \ ,
\end{aligned}
\end{equation}
where we used (\ref{sigmaU_iso}) together with (\ref{sigma_U_char_U}) and (\ref{sigma_char_U}), as well as (\ref{X_characters}). Also recall the notation $[a]:=a \ \text{mod} \ 1 $. Similarly, we find
\begin{equation}
\begin{aligned}
        \text{ch}\bigl[\bigoplus_{\substack{r,w\in \mathbb{Z}: \\ r = w+1 \, \text{mod}\,2 }}  \sigma^{w}\sigma^{r}_U \bigl(\mathcal{R}^+ \bigl)\bigl] 
        & = \sum_{U \in \frac{1}{2}\mathbb{Z}} \frac{q^{-U^2}}{\eta(\tau)} (-1)^{2U} \, \text{sch}\bigl[\sigma^{2[U]}\bigl(\mathbb{L} \bigl) \bigl] \ .
\end{aligned}
\end{equation}
Now, recall that a modular invariant for $\mathfrak{u}(1)_{-1/2}$ is given by
\begin{equation}\label{u1_mod_inv}
  \sum_{\substack{U,U' \in \frac{1}{2}\mathbb{Z} \, :\\ U = U' \, \text{mod}\,1}} \frac{q^{-U^2}}{\eta(\tau)} \frac{\bar{q}^{-U'^2}}{\eta(\bar{\tau})} \ ,
\end{equation}
and the modular $S$-matrix for the modules $(\mathbb{L}, \sigma\bigl(\mathbb{L}\bigl))$ is given by (\ref{su2_1_S_matr}). At this point, we combine (\ref{proposal}) together with the analogous contribution for the other boundary conditions using (\ref{tilde_char}). Then, the full free field partition function decomposes in the product of (\ref{u1_mod_inv}) with the partition function of (\ref{su2_1_spectrum}), and hence it is modular invariant.

\section{Conclusions}\label{sec:concl}
\setcounter{equation}{0}

In this paper we elucidated some aspects of the WZW model on $\mathfrak{su}(2)_{-1}$. This model has a free field realisation in terms of four symplectic bosons, for whose free field representations we found an interesting decomposition in terms of $\mathfrak{su}(2)_{-1}$ modules. We gave continuous modular invariants for the $\mathfrak{su}(2)_{-1}$ theory involving only discrete (\ref{discr_spectra}) and only continuous (\ref{cont_diag_spectrum}) $\mathfrak{su}(2)$ representations, which are analogous to those found in \cite{Maldacena:2000hw}. 
More interestingly, we found two invariant spectra containing only a discrete subset of representations. We understood them as arising through a coset construction from invariants of $\mathfrak{psu}(2|2)_1$ found in \cite{Gaberdiel:2023nhb}. We also gave a possible interpretation of them as modular invariants of simple current (super)algebra extensions of $\mathfrak{su}(2)_{-1}$. It is this latter picture that we find intriguing: one invariant carries the same representation under the modular group as $\mathfrak{su}(2)_1$, but the extension seems to be supersymmetric in nature. The second invariant seems to be associated to a (bosonic) rational CFT with eight irreducible representations. It would be interesting to work out the details of this correspondence and understand the vertex operator (super)algebras arising as extensions of $\mathfrak{su}(2)_{-1}$. More generally, it is intriguing that a WZW model at negative level can be expressed in terms of simple current extensions; this has been observed e.g. at fractional level in \cite{Creutzig:2012sd}. 

It is also worth commenting on the free field theory of symplectic bosons compared to that of fermions. One of the crucial differences is that spectral flow automorphisms have infinite orbit, which complicates the analysis of the characters and their modular properties because of convergence issues. Nevertheless, we managed to give a free field modular invariant for four symplectic bosons, which analogously to the theory of fermions on the torus, it requires to include different boundary conditions of the bosonic fields.

This work was mainly motivated by the exploration of supersymmetric WZW models for the description of tensionless string theory on ${\rm AdS}$ spaces \cite{Gaberdiel:2023nhb}. Independently from that, we hope that these computations can help in the better understanding of non-unitary 2d CFTs with negative central charge, as well as logarithmic CFTs, provided by $\mathfrak{su}(2)_{-1}$.

\section*{Acknowledgements}

This paper collects some results based on the continuation of my Master thesis. I am very grateful to Matthias Gaberdiel for guiding me through my Master's project and taking the time every week to meet and talk with me. I thank David Ridout for insightful conversations. Lastly, I also thank Martín Lagares and Prashanth Raman for useful comments on the draft.

\clearpage

\appendix

\section{Different R sectors}\label{app_different_R}
\setcounter{equation}{0}
\label{appendix_shortening}
\renewcommand{\theequation}{A.\arabic{equation}}

As mentioned in Footnote \ref{footn}, there is freedom in defining the action of the symplectic boson zero modes in the R sector, and different choices are related by the conjugation automorphism (\ref{conj_aut}). It turns out that different choices are relevant only when considering the R sector with occupation numbers $m_1,m_2 \in \tfrac{1}{2}\mathbb{N}_0$. Previously we defined the free field module $\mathcal{R}^{+}$ by (\ref{sympl_bos_zero}) with $\delta_1=\delta_2=0$. Here we explicitely define the other R sectors and focus on their subrepresentation determined by $m_i \in \tfrac{1}{2}\mathbb{N}_0$. It turns out that different choices for the action of the zero modes give rise to different types of $\mathfrak{su}(2)_{-1}$ representations: in $\mathcal{R}^{+}$ we found those associated to finite-dimensional $\mathfrak{su}(2)$ representations, and now we recover also the highest/lowest weight discrete ones. Moreover, we show that these sectors are related by different spectral flow actions.

We start from the (fully conjugate with respect to (\ref{sympl_bos_zero})) R sector 
\begin{equation}
\label{R-}
     \begin{aligned}[t]
     \lambda^{1}_{0} \, |m_{1}, m_{2} \rangle & = | m_{1} + \tfrac{1}{2}, m_{2} \rangle \ , \\
    \lambda^{2}_{0} \, |m_{1}, m_{2} \rangle & =  | m_{1} , m_{2} + \tfrac{1}{2} \rangle \ ,
    \end{aligned}
    \qquad \qquad
    \begin{aligned}[t]
    (\mu^{\dagger}_{1})_{0} \, |m_{1}, m_{2} \rangle & = -2m_{1} \, | m_{1} - \tfrac{1}{2}, m_{2} \rangle \ , \\
    (\mu^{\dagger}_{2})_{0} \, |m_{1}, m_{2} \rangle & = -2  m_{2} \, | m_{1} , m_{2} - \tfrac{1}{2} \rangle \ ,
    \end{aligned}
\end{equation}
from which we compute
\begin{equation}
    \begin{aligned}
     J^{3}_{0} \, | m_{1}, m_{2} \rangle &= (m_{1} - m_{2}) \, | m_{1}, m_{2} \rangle \ , \\
     J^{+}_{0} \, | m_{1}, m_{2} \rangle &= -2  m_{2} \, | m_{1} + \tfrac{1}{2}, m_{2} - \tfrac{1}{2} \rangle \ , \\
     J^{-}_{0} \, | m_{1}, m_{2} \rangle &= -2  m_{1} \, | m_{1} - \tfrac{1}{2}, m_{2} + \tfrac{1}{2} \rangle \ , \\
     U_{0} \, | m_{1}, m_{2} \rangle &= -(m_{1} + m_{2} + \tfrac{1}{2}) \, | m_{1}, m_{2} \rangle \ , \\
    \end{aligned}
\end{equation}
and the $\mathfrak{su}(2)$-Casimir
\begin{equation}
    C^{\mathfrak{su}(2)} = j  (j + 1) = (m_{1} + m_{2}) \, (m_{1} + m_{2} + 1) \ ,
\end{equation}
thus the associated spin is again $j = m_{1} + m_{2}$. We denote by $\mathcal{R}^{-}$ the full R sector defined by (\ref{R-}) with $m_{1},m_{2} \in \tfrac{1}{2}\mathbb{N}_0$. Then $\mathcal{R}^{-}$ is conjugate to $\mathcal{R}^{+}$ in the sense that
\begin{equation}
    \text{ch} \bigl[\mathcal{R}^+\bigl](t,\mu;\tau) = \text{ch} \bigl[\mathcal{R}^-\bigl](-t,-\mu;\tau) \,,
\end{equation}
thus from (\ref{R_decomposition}) it follows that
\begin{equation}\label{R__decomp}
    \mathcal{R}^{-} \cong \bigoplus_{U \in \frac{1}{2}\mathbb{Z}} \mathcal{R}^{-}_{U}~~~\text{with}~~~ \mathcal{R}^{-}_{U} \cong \mathcal{F}_{U} \otimes \bigoplus_{j \in \mathbb{N}+|U+\frac{1}{2}|}  \mathcal{H}_{j}  \ .
\end{equation}
From (\ref{R__decomp}) it readily follows that
\begin{equation}\label{sigma_U__}
    \sigma_U \bigl( \mathcal{V} \bigl) \cong \mathcal{R}^{-} \ .
\end{equation}

Another possible R sector is defined by
\begin{equation}
\label{R+_discrete}
     \begin{aligned}[t]
     \lambda^{1}_{0} \, |m_{1}, m_{2} \rangle & = 2  m_{1} \, | m_{1} - \tfrac{1}{2}, m_{2} \rangle \ , \\
    \lambda^{2}_{0} \, |m_{1}, m_{2} \rangle & =  | m_{1} , m_{2} + \tfrac{1}{2} \rangle \ ,
    \end{aligned}
    \qquad \qquad
    \begin{aligned}[t]
    (\mu^{\dagger}_{1})_{0} \, |m_{1}, m_{2} \rangle & = | m_{1} + \tfrac{1}{2}, m_{2} \rangle \ , \\
    (\mu^{\dagger}_{2})_{0} \, |m_{1}, m_{2} \rangle & = - 2  m_{2} \, | m_{1} , m_{2} - \tfrac{1}{2} \rangle \ .
    \end{aligned}
\end{equation}
from which we compute
\begin{equation}
    \begin{aligned}
     J^{3}_{0} \, | m_{1}, m_{2} \rangle &= -(m_{1} + m_{2} + \tfrac{1}{2}) \,  | m_{1}, m_{2} \rangle \ , \\
     J^{+}_{0} \, | m_{1}, m_{2} \rangle &= -4  m_{1}  m_{2} \, | m_{1} - \tfrac{1}{2}, m_{2} - \tfrac{1}{2} \rangle \ , \\
     J^{-}_{0} \, | m_{1}, m_{2} \rangle &= | m_{1} + \tfrac{1}{2}, m_{2} + \tfrac{1}{2} \rangle \ , \\
     U_{0} \, | m_{1}, m_{2} \rangle &= (m_{1} - m_{2} ) \, | m_{1}, m_{2} \rangle \ , \\
    \end{aligned}
\end{equation}
and the $\mathfrak{su}(2)$-Casimir
\begin{equation}
    C^{\mathfrak{su}(2)} = j  (j + 1) = (m_{1} - m_{2} - \tfrac{1}{2})  (m_{1} - m_{2} + \tfrac{1}{2}) \ ,
\end{equation}
with spin $j = m_{1} - m_{2} - \tfrac{1}{2}$. We denote the R sector defined by (\ref{R+_discrete}) with $m_{1},m_{2} \in \tfrac{1}{2}\mathbb{N}_0$ by~$\mathcal{D}^{+}$.
A result similar to (\ref{R_character_decomposition}) and (\ref{R__decomp}) holds for $\mathcal{D}^+$, which contains highest weight discrete representations of $\mathfrak{su}(2)$ to spin $j \in  \tfrac{1}{2}\mathbb{Z}_{<0}$. Indeed, its character is given by 
\begin{equation}
\label{R_character_discrete}
    \begin{aligned}
        \text{ch}\bigl[\mathcal{D}^+\bigl](t,\mu;\tau) &= q^{-\frac{1}{6}} \sum_{m_{1},m_{2} \in \frac{1}{2}\mathbb{N}} x^{-m_{1}-m_{2} - \frac{1}{2}}y^{m_1 - m_{2} } \prod_{n=1}^{\infty} \prod_{a,b = \pm \frac{1}{2}} \frac{1}{1 - x^{a}\, y^{b} \, q^{n} } \\
        &= \frac{\eta(\tau)^2}{\vartheta_{1}(\frac{t+\mu}{2};\tau)\vartheta_{1}(\frac{t-\mu}{2};\tau)}  \ ,
    \end{aligned}
\end{equation}
valid for $1<|x|<|q|^{-2}$ and $|q|<|y|<|q|^{-1}$. Then, by denoting with $\mathcal{D}^+\bigl|_U \subset \mathcal{D}^+$ the subrepresentation to fixed $U_0=U \in \tfrac{1}{2}\mathbb{Z}$, we have
\begin{equation}\label{R_character_decomposition_discrete}
    \text{ch}\bigl[\mathcal{D}^+\bigl|_U\bigl](t;\tau)  = \frac{q^{-U^2}}{\eta(\tau)} \sum_{j \in \mathbb{N} + |U|} \text{ch}\bigl[\mathcal{D}^{+}_{-j-\frac{1}{2}}\bigl](t;\tau) \ ,
\end{equation}
which can be also proven by applying (\ref{denominator_identity}) with the substitutions $u = x^{-\frac{1}{2}}y^{\frac{1}{2}}$ and $v = x^{-\frac{1}{2}}y^{-\frac{1}{2}}$; note that this changes the convergence region to ${1 < |x|^{\frac{1}{2}}|y|^{-\frac{1}{2}},} |x|^{\frac{1}{2}}|y|^{\frac{1}{2}} < |q|^{-1}$, which needs to be taken into account in order to obtain
\begin{equation}
\begin{aligned}
        \sum_{m,n=0}^{\infty}x^{-\frac{n+m+1}{2}}y^{\frac{n-m-1}{2}} \prod_{n=1}^{\infty} \frac{x^{\frac{1}{2}}(1-q^{n})^2(1- x q^{n})(1-x^{-1} q^{n-1})}{(1- x^{\frac{1}{2}}y^{-\frac{1}{2}}q^{n})(1-x^{-\frac{1}{2}}y^{\frac{1}{2}}q^{n})(1-x^{\frac{1}{2}}y^{\frac{1}{2}} q^{n})(1- x^{-\frac{1}{2}}y^{-\frac{1}{2}}q^{n})}  \\
    = \Bigg(\sum_{m,n=0}^{\infty}-\sum_{m,n=-1}^{-\infty} \Bigg) x^{-\frac{n+m}{2}}y^{\frac{m-n-1}{2}}q^{mn} \ .
\end{aligned}
\end{equation}
Then, (\ref{R_character_decomposition_discrete}) is equivalent to
\begin{equation}
\label{R_decomposition_discrete}
    \mathcal{D}^+ \cong \bigoplus_{U \in \frac{1}{2}\mathbb{Z}} \mathcal{D}^+\bigl|_U~~~\text{with}~~~\mathcal{D}^+\bigl|_U \cong \mathcal{F}_{U}  \otimes \bigoplus_{j \in \mathbb{N} + |U| }  \mathcal{D}^{+}_{-j-\frac{1}{2}}  \ ,
\end{equation}
from which one infers
\begin{equation}\label{sigma_iso_fields+}
    \sigma\bigl( \mathcal{R}^+ \bigl) \cong \mathcal{D}^+ \ .
\end{equation}

The last R sector we consider is (the fully conjugate one with respect to $\mathcal{D}^{+}$)
\begin{equation}
\label{R-_discrete}
     \begin{aligned}[t]
     \lambda^{1}_{0} \, |m_{1}, m_{2} \rangle & =   | m_{1} + \tfrac{1}{2}, m_{2} \rangle \ , \\
    \lambda^{2}_{0} \, |m_{1}, m_{2} \rangle & = 2  m_{2} \, | m_{1} , m_{2} - \tfrac{1}{2} \rangle \ ,
    \end{aligned}
    \qquad \qquad
    \begin{aligned}[t]
    (\mu^{\dagger}_{1})_{0} \, |m_{1}, m_{2} \rangle & = -2 m_{1}  | m_{1} - \tfrac{1}{2}, m_{2} \rangle \ , \\
    (\mu^{\dagger}_{2})_{0} \, |m_{1}, m_{2} \rangle & =  | m_{1} , m_{2} + \tfrac{1}{2} \rangle \ ,
    \end{aligned}
\end{equation}
from which we compute
\begin{equation}
    \begin{aligned}
     J^{3}_{0} \, | m_{1}, m_{2} \rangle &= (m_{1} + m_{2} + \tfrac{1}{2}) \,  | m_{1}, m_{2} \rangle \ , \\
     J^{+}_{0} \, | m_{1}, m_{2} \rangle &= | m_{1} + \tfrac{1}{2}, m_{2} + \tfrac{1}{2} \rangle \ , \\
     J^{-}_{0} \, | m_{1}, m_{2} \rangle &=  -4  m_{1}  m_{2} \, | m_{1} - \tfrac{1}{2}, m_{2} - \tfrac{1}{2} \rangle \ , \\
     U_{0} \, | m_{1}, m_{2} \rangle &= (m_{2} - m_{1} ) \, | m_{1}, m_{2} \rangle \ , \\
    \end{aligned}
\end{equation}
and the $\mathfrak{su}(2)$-Casimir
\begin{equation}
    C^{\mathfrak{su}(2)} = j  (j - 1) = (m_{2} - m_{1} + \tfrac{1}{2})  (m_{2} - m_{1} - \tfrac{1}{2}) \ ,
\end{equation}
thus the spin is $j = m_{2} - m_{1} + \tfrac{1}{2}$. We denote its subsector $m_{1},m_{2} \in \tfrac{1}{2}\mathbb{N}_0$ by $\mathcal{D}^{-}$. Notice that $\mathcal{D}^{-}$ is conjugate to $\mathcal{D}^{+}$ in the sense that  
\begin{equation}
    \text{ch}\bigl[\mathcal{D}^+\bigl](t,\mu;\tau) = \text{ch}\bigl[\mathcal{D}^-\bigl](-t,-\mu;\tau) \ ,
\end{equation}
which also changes the convergence region to $|q|^{2}<|x|<1$ and $|q|<|y|<|q|^{-1}$ in (\ref{R_character_discrete}) and implies together with (\ref{R_character_decomposition_discrete}) that
\begin{equation}\label{R-_char_decomposition}
    \text{ch}\bigl[\mathcal{D}^{-} \bigl|_U \bigl](t;\tau) = \frac{q^{-U^2}}{\eta(\tau)} \sum_{j \in \mathbb{N} + |U|} \text{ch}\bigl[\mathcal{D}^{-}_{j + \frac{1}{2}}\bigl](t;\tau) \ ,
\end{equation}
or equivalently,
\begin{equation}\label{D__decomp}
    \mathcal{D}^- \cong \bigoplus_{U \in \frac{1}{2}\mathbb{Z}} \mathcal{D}^-\bigl|_U ~~\text{with}~~ \mathcal{D}^-\bigl|_U \cong \mathcal{F}_{U} \otimes \bigoplus_{j \in \mathbb{N}+|U|} \mathcal{D}^{-}_{j+\frac{1}{2}}  \ .
\end{equation}
From (\ref{D__decomp}) one deduces 
\begin{equation}\label{sigma_iso_fields_}
    \sigma^{-1} \bigl( \mathcal{R}^{+} \bigl) \cong \mathcal{D}^{-} \ .
\end{equation}

Putting these results together, we can decompose the character of $\mathcal{R}_U$ in terms of those of $\mathcal{R}^+_{U}$ and spectrally flowed images of it. More precisely, we start from the character of the full R sector $\mathcal{R}$, which is defined by (\ref{char_sympl_bos}) with $\delta_1 = \delta_2 = 0$, we decompose the summation over $m_{1}, m_{2} \in \tfrac{1}{2}\mathbb{Z}$ in the four sectors $m_i \geq 0$, $m_i < 0$, $m_1 \geq 0$ and $m_2 < 0$, $m_1 < 0$ and $m_2 \geq 0$, and extract the coefficient to $y^U$. We find 
\begin{equation}\label{R_char_sympl_bos_decompos}
\begin{aligned}
        \text{ch}\bigl[\mathcal{R}_U \bigl] &= \text{ch}\bigl[\mathcal{R}^+_U \bigl] + \, \text{ch}\bigl[\mathcal{R}^-_{U} \bigl] + \, \text{ch}\bigl[\mathcal{D}^+\bigl|_U \bigl] + \,  \text{ch}\bigl[\mathcal{D}^-\bigl|_{U} \bigl] \\
        & = \text{ch}\bigl[\mathcal{R}^+_U \bigl] + \, \text{ch}\bigl[\sigma_U^{2}\bigl(\mathcal{R}^+_{U-1}\bigl) \bigl] + \, \text{ch}\bigl[\sigma \bigl(\mathcal{R}^+_U\bigl) \bigl] +\,  \text{ch}\bigl[\sigma^{-1}\bigl(\mathcal{R}^+_{U} \bigl) \bigl] \,,
\end{aligned}
\end{equation}
where in the second equality we used the isomorphisms (\ref{sigmaU_iso}), (\ref{sigma_U__}), (\ref{sigma_iso_fields+}) and (\ref{sigma_iso_fields_}). Looking at the various decomposition in $\mathfrak{su}(2)_{-1}$ irreducibles, as for instance (\ref{R_decomposition}), (\ref{R_char_sympl_bos_decompos}) incorporates (\ref{continuous_decomp}). We also mention that (\ref{R_char_sympl_bos_decompos}) can be used in \cite[(3.28-29)]{Gaberdiel:2023nhb} to show the composition series of the continuous indecomposable modules of $\mathfrak{u}(2|2)_1$.

\section{Theta functions}
\label{theta_functions}

\setcounter{equation}{0}
\renewcommand{\theequation}{B.\arabic{equation}}

We define 
\begin{equation}
\begin{aligned}
       \vartheta \begin{bmatrix} \alpha \\ \beta
               \end{bmatrix} (z;\tau) & := \sum_{n\in \mathbb{Z}} {\rm e}^{\pi i (n+\alpha)^{2}\tau + 2\pi i (n+\alpha )(z + \beta)} \\
               &= {\rm e}^{2 \pi i  \alpha(z + \beta) } q^{\frac{\alpha^{2}}{2}} \prod_{n = 1}^{\infty}(1-q^{n})(1+q^{n + \alpha - \frac{1}{2}} {\rm e}^{2 \pi i (z + \beta )})(1 + q^{n- \alpha -\frac{1}{2}} {\rm e}^{-2 \pi i (z + \beta)}) \ ,
\end{aligned}
\end{equation}
where the second equality holds by applying the \textit{Jacobi triple product}. The Jacobi theta functions are then

\begin{equation}
    \vartheta_{1} := \vartheta \begin{bmatrix} \frac{1}{2} \\ \frac{1}{2}
               \end{bmatrix} \ ,~~~~ \vartheta_{2} := \vartheta \begin{bmatrix} \frac{1}{2} \\ 0
               \end{bmatrix} \ ,~~~~ \vartheta_{3} := \vartheta \begin{bmatrix} 0 \\ 0
               \end{bmatrix} \ ,~~~~ \vartheta_{4} := \vartheta \begin{bmatrix} 0 \\ \frac{1}{2}
               \end{bmatrix} \ .
\end{equation}
These functions obey the following addition rules:
\begin{equation}\label{addition_rules}
\begin{aligned}
        \vartheta_{1}(\tfrac{z+t}{2};\tau)\vartheta_{1}(\tfrac{z-t}{2};\tau) &= \vartheta_{2}(z;2\tau) \vartheta_{3}(t;2\tau)- \vartheta_{3}(z;2\tau) \vartheta_{2}(t;2\tau) \ , \\        
        \vartheta_{2}(\tfrac{z+t}{2};\tau)\vartheta_{2}(\tfrac{z-t}{2};\tau) &= \vartheta_{2}(z;2\tau) \vartheta_{3}(t;2\tau) + \vartheta_{3}(z;2\tau) \vartheta_{2}(t;2\tau) \ , \\        
        \vartheta_{3}(\tfrac{z+t}{2};\tau)\vartheta_{3}(\tfrac{z-t}{2};\tau) &= \vartheta_{3}(z;2\tau) \vartheta_{3}(t;2\tau)- \vartheta_{2}(z;2\tau) \vartheta_{2}(t;2\tau) \ , \\       
        \vartheta_{4}(\tfrac{z+t}{2};\tau)\vartheta_{4}(\tfrac{z-t}{2};\tau) &= \vartheta_{3}(z;2\tau) \vartheta_{3}(t;2\tau)- \vartheta_{2}(z;2\tau) \vartheta_{2}(t;2\tau) \ , \\ 
\end{aligned}
\end{equation}
and quasi-periodicity relations, such as
\begin{equation}
\label{theta_function_periodicity1}
\begin{aligned}
    \vartheta_{i}(z + w \tau;\tau) &= {\rm e}^{- 2 \pi i w z}q^{\frac{-w^2}{2}} \vartheta_{i}(z;\tau) ~~~~~~~~~~~~ i =2,3 \ , \\
    \vartheta_{i}(z + w \tau;\tau) &= (-1)^{w}{\rm e}^{- 2 \pi i w z}q^{\frac{-w^2}{2}} \vartheta_{i}(z;\tau) ~~~~ i =1,4 \ ,
\end{aligned}
\end{equation}
for every $w \in \mathbb{Z}$, and 
\begin{equation}
\label{theta_function_periodicity2}
\begin{aligned}
   \vartheta_{2}(z + w\tau ; 2\tau) &= {\rm e}^{-  \pi i w z} q^{-\frac{w^2}{4}} \begin{cases}
        \vartheta_{2}(z;2\tau)~~~~ \forall \, w \in 2\mathbb{Z} \,\\
        \vartheta_{3}(z;2\tau)~~~~ \forall \, w \in 2\mathbb{Z} + 1 \ ,
    \end{cases} \\
    \vartheta_{3}(z + w \tau ; 2\tau) &= {\rm e}^{-  \pi i w z} q^{-\frac{w^2}{4}} \begin{cases}
        \vartheta_{3}(z;2\tau)~~~~ \forall \, w \in 2\mathbb{Z} \,\\
        \vartheta_{2}(z;2\tau)~~~~ \forall \, w \in 2\mathbb{Z} + 1 \ .
    \end{cases} 
\end{aligned} 
\end{equation}
We also need the modular transformations
\begin{equation}\label{theta_mod_tr}
    \begin{aligned}
        \vartheta_{1}(\tfrac{z}{\tau}; -\tfrac{1}{\tau}) &= - i \sqrt{-i \tau}\,  {\rm e}^{\frac{\pi i z^{2}}{\tau}} \vartheta_{1}(z;\tau) \ , \\
        \vartheta_{2}(\tfrac{z}{\tau}; -\tfrac{1}{\tau}) &=  \sqrt{-i \tau}\,  {\rm e}^{\frac{\pi i z^{2}}{\tau}} \vartheta_{4}(z;\tau) \ , \\
        \vartheta_{3}(\tfrac{z}{\tau}; -\tfrac{1}{\tau}) &=  \sqrt{-i \tau}\,  {\rm e}^{\frac{\pi i z^{2}}{\tau}} \vartheta_{3}(z;\tau) \ , \\
        \vartheta_{4}(\tfrac{z}{\tau}; -\tfrac{1}{\tau}) &=  \sqrt{-i \tau}\,  {\rm e}^{\frac{\pi i z^{2}}{\tau}} \vartheta_{2}(z;\tau) \ , \\
    \end{aligned}
\end{equation}
and 
\begin{equation}\label{theta2_mod_tr}
    \begin{aligned}
        \vartheta_{2}(\tfrac{z}{\tau}; -\tfrac{2}{\tau}) &=  \sqrt{-\tfrac{i \tau}{2}}\,  {\rm e}^{\frac{\pi i z^{2}}{2 \tau}} \big( \vartheta_{3}(z;2\tau) -  \vartheta_{2}(z;2\tau) \big) \ , \\
        \vartheta_{3}(\tfrac{z}{\tau}; -\tfrac{2}{\tau}) &=  \sqrt{-\tfrac{i \tau}{2}}\,  {\rm e}^{\frac{\pi i z^{2}}{2 \tau}} \big( \vartheta_{3}(z;2\tau) +  \vartheta_{2}(z;2\tau) \big) \ . \\
    \end{aligned}
\end{equation}
We also make use of the Dedekind eta function, which is defined by
\begin{equation}
    \eta(\tau) = q^{\frac{1}{24}} \prod_{n=1}^{\infty}(1-q^{n}) \ ,
\end{equation}
and it transform under the modular group as
\begin{equation}\label{eta_modular_tr}
    \begin{aligned}
        \eta(\tau + 1) &= {\rm e}^{\frac{\pi i }{12}} \, \eta(\tau) \ , \\
        \eta(-\tfrac{1}{\tau}) &= \sqrt{-i \tau} \, \eta(\tau) \ . \\
    \end{aligned}
\end{equation}



\begin{thebibliography}{99}

\bibitem{Maldacena:2000hw}
J.~M.~Maldacena and H.~Ooguri,
``Strings in ${\rm AdS}_3$ and $\SL(2,\mathbb{R})$ WZW model 1.: The Spectrum,''
J. Math. Phys. \textbf{42} (2001) 2929-2960
{\tt[\href{https://arxiv.org/abs/hep-th/0001053}{arXiv:hep-th/0001053 [hep-th]}]{}}


\bibitem{Eberhardt:2018ouy}
L.~Eberhardt, M.R.~Gaberdiel and R.~Gopakumar,
``The Worldsheet Dual of the Symmetric Product CFT,''
JHEP \textbf{04} (2019) 103
{\tt [\href{https://arxiv.org/abs/1812.01007}{arXiv:1812.01007 [hep-th]}]}.

\bibitem{Eberhardt:2019ywk}
L.~Eberhardt, M.R.~Gaberdiel and R.~Gopakumar,
``Deriving the AdS$_{3}$/CFT$_{2}$ correspondence,''
JHEP \textbf{02} (2020) 136
{\tt[\href{https://arxiv.org/abs/1911.00378}{arXiv:1911.00378 [hep-th]}]{}}.


\bibitem{Gaberdiel:2021qbb}
M.R.~Gaberdiel and R.~Gopakumar,
``String Dual to Free N=4 Supersymmetric Yang-Mills Theory,''
Phys.\ Rev.\ Lett.\ \textbf{127} (2021) 131601
{\tt [\href{https://arxiv.org/abs/2104.08263}{arXiv:2104.08263 [hep-th]}]}.

\bibitem{Gaberdiel:2021jrv}
M.R.~Gaberdiel and R.~Gopakumar,
``The worldsheet dual of free super Yang-Mills in 4D,''
JHEP \textbf{11} (2021) 129
{\tt [\href{https://arxiv.org/abs/2105.10496}{arXiv:2105.10496 [hep-th]}]}.


\bibitem{Beisert:2005tm}
N.~Beisert,
``The SU($2|2$) dynamic S-matrix,''
Adv.\ Theor.\ Math.\ Phys.\ \textbf{12} (2008) 945-979
{\tt[\href{https://arxiv.org/abs/hep-th/0511082}{arXiv:hep-th/0511082 [hep-th]}]{}}.



\bibitem{Gaberdiel:2023nhb}
M.~R.~Gaberdiel and E.~Mazzucchelli,
``The $\mathfrak{u}(2|2)_1$ WZW model,''
 J.Phys.A \textbf{57} (2024) 17, 175401
{\tt
[\href{https://arxiv.org/abs/2312.03135}{arXiv:2312.03135[hep-th]}]}.

\bibitem{Conf_emb}
D.~Adamovic, P.~M\"oseneder, P.~Papi and O.~Perse,
``Conformal embeddings in affine vertex superalgebras,''
Adv.\ Math.\ {\bf 360} (2020) 106918 
{\tt[\href{https://arxiv.org/abs/1903.03794}{arXiv:1903.03794 [math.RT]}]{}}.


\bibitem{Ridout:2008nh}
D.~Ridout,
``$\hat{\mathfrak{sl}}(2)_{-1/2}$: A Case Study,''
Nucl.\ Phys.\ B \textbf{814} (2009) 485-521
{\tt[\href{https://arxiv.org/abs/0810.3532}{arXiv:0810.3532 [hep-th]}]}.


\bibitem{Flohr:2001zs}
M.~Flohr,
``Bits and pieces in logarithmic conformal field theory,''
Int.\ J.\ Mod.\ Phys.\ A \textbf{18} (2003) 4497-4592
{\tt[\href{https://arxiv.org/abs/hep-th/0111228}{arXiv:hep-th/0111228 [hep-th]}]{}}.

\bibitem{Gaberdiel:2001tr}
M.R.~Gaberdiel,
``An Algebraic approach to logarithmic conformal field theory,''
Int.\ J.\ Mod.\ Phys. A \textbf{18} (2003) 4593-4638
{\tt[\href{https://arxiv.org/abs/hep-th/0111260}{arXiv:hep-th/0111260 [hep-th]}]{}}.

\bibitem{Creutzig:2013hma}
T.~Creutzig and D.~Ridout,
``Logarithmic Conformal Field Theory: Beyond an Introduction,''
J.\ Phys.\ A \textbf{46} (2013) 4006
{\tt[\href{https://arxiv.org/abs/1303.0847}{arXiv:1303.0847 [hep-th]}]{}}.


\bibitem{Gurdogan:2015csr}
\"O.~G\"urdo\u{g}an,V.~Kazakov,
``New Integrable 4D Quantum Field Theories from Strongly Deformed Planar $\mathcal N = $ 4 Supersymmetric Yang-Mills Theory,''
 JHEP \textbf{20} (2016) 117
{\tt
[\href{https://arxiv.org/abs/1512.06704}{arXiv:1512.06704[hep-th]}]}.

\bibitem{Kazakov:2022dbd}
V.~Kazakov, E.~Olivucci, 
``The loom for general fishnet CFTs,''
 JHEP \textbf{06} (2023) 041
{\tt
[\href{https://arxiv.org/abs/2212.09732}{arXiv:2212.09732[hep-th]}]}.




\bibitem{Schomerus:2005bf}
V.~Schomerus and H.~Saleur,
``The GL($1|1$) WZW model: From supergeometry to logarithmic CFT,''
Nucl.\ Phys.\ B \textbf{734} (2006) 221-245
{\tt[\href{https://arxiv.org/abs/hep-th/0510032}{arXiv:hep-th/0510032 [hep-th]}]{}}.

\bibitem{Gotz:2006qp}
G.~Gotz, T.~Quella and V.~Schomerus,
``The WZNW model on PSU($1,1|2$),''
JHEP \textbf{03} (2007) 003
{\tt{\href{https://arxiv.org/abs/hep-th/0510032}{arXiv:hep-th/0610070 [hep-th]}}}.

\bibitem{Saleur:2006tf}
H.~Saleur and V.~Schomerus,
``On the SU($2|1$) WZW model and its statistical mechanics applications,''
Nucl.\ Phys.\ B \textbf{775} (2007) 312-340
{\tt[\href{https://arxiv.org/abs/hep-th/0611147}{arXiv:hep-th/0611147 [hep-th]}]{}}.



\bibitem{Flohr:1995ea}
M.A.I.~Flohr,
``On modular invariant partition functions of conformal field theories with logarithmic operators,''
Int.\ J.\ Mod.\ Phys.\ A \textbf{11} (1996) 4147-4172
{\tt[\href{https://arxiv.org/abs/hep-th/9509166}{arXiv:hep-th/9509166 [hep-th]}]{}}.

\bibitem{Creutzig:2012sd}
T.~Creutzig and D.~Ridout,
``Modular Data and Verlinde Formulae for Fractional Level WZW Models I,''
Nucl.\ Phys.\ B \textbf{865} (2012) 83-114
{\tt[\href{https://arxiv.org/abs/1205.6513}{arXiv:1205.6513 [hep-th]}]}.

\bibitem{Creutzig:2013yca}
T.~Creutzig and D.~Ridout,
``Modular Data and Verlinde Formulae for Fractional Level WZW Models II,''
Nucl.\ Phys.\ B \textbf{875} (2013) 423-458
{\tt[\href{https://arxiv.org/abs/1306.4388}{arXiv:1306.4388 [hep-th]}}.




\bibitem{Ridout:2010jk}
D.~Ridout,
``Fusion in Fractional Level $\hat{\mathfrak{sl}}$(2)-Theories with $k = - \frac{1}{2}$,''
Nucl.\ Phys.\ B \textbf{848} (2011) 216-250
{\tt[\href{https://arxiv.org/abs/1012.2905}{arXiv:1012.2905 [hep-th]}]}.


\bibitem{Gaberdiel:2001ny}
M.R.~Gaberdiel,
``Fusion rules and logarithmic representations of a WZW model at fractional level,''
Nucl.\ Phys.\ B \textbf{618} (2001) 407-436
{\tt[\href{https://arxiv.org/abs/hep-th/0105046}{arXiv:hep-th/0105046 [hep-th]}]{}}.


\bibitem{Goddard:1987td}
P.~Goddard, D.I.~Olive and G.~Waterson,
``Superalgebras, Symplectic Bosons and the Sugawara Construction,''
Commun.\ Math.\ Phys.\ \textbf{112} (1987) 591.


\bibitem{Gaberdiel_2018}
M.R.~Gaberdiel and R.~Gopakumar,
``Tensionless string spectra on ${\rm AdS}_3$,''
JHEP \textbf{05} (2018) 085
{\tt [\href{https://arxiv.org/abs/1803.04423}{arXiv:1803.04423 [hep-th]}]}.



\bibitem{Kac}
V.~G.~Kac and D.~A.~Kazhdan
``Structure of representations with highest weight of infinite-dimensional Lie algebras,''
Advances \ in \ Mathematics \ \textbf{34} (1979) 97-108





\bibitem{Lesage:2003kn}
F.~Lesage, P.~Mathieu, J.~Rasmussen and H.~Saleur,
``Logarithmic lift of the affine $\mathfrak{su}(2)_{-1/2}$ model,''
Nucl. Phys. B \textbf{686} (2004), 313-346
{\tt
[\href{https://arxiv.org/abs/hep-th/0311039}{arXiv:hep-th/0311039  [hep-th]}]}.



\bibitem{Kac:1994kn}
V.~G.~Kac and M.~Wakimoto,
``Integrable highest weight modules over affine superalgebras and number theory,''
{\tt[\href{https://arxiv.org/abs/hep-th/9407057}{arXiv:hep-th/9407057 [hep-th]}]{}}
























\bibitem{Metsaev:1998it}
R.R.~Metsaev and A.A.~Tseytlin,
``Type IIB superstring action in AdS$_5 \times {\rm S}^5$ background,''
Nucl.\ Phys.\ B \textbf{533} (1998) 109-126
{\tt[\href{https://arxiv.org/abs/hep-th/9805028}{arXiv:hep-th/9805028 [hep-th]}]{}}.




\bibitem{Gaberdiel:2022als}
M.~R.~Gaberdiel, K.~Naderi and V.~Sriprachyakul,
``The free field realisation of the BVW string,''
JHEP \textbf{08} (2022) 274
{\tt
[\href{https://arxiv.org/abs/2202.11392}{arXiv:2202.11392 [hep-th]}]}.


\bibitem{Gaberdiel:2011vf}
M.~R.~Gaberdiel and S.~Gerigk,
``The massless string spectrum on ${\rm AdS}_3\times { \rm S}^{3}$ from the supergroup,''
JHEP \textbf{10} (2011) 045
{\tt[\href{https://arxiv.org/abs/1107.2660}{arXiv:1107.2660 [hep-th]}]}



\bibitem{Dei:2020zui}
A.~Dei, M.R.~Gaberdiel, R.~Gopakumar and B.~Knighton,
``Free field world-sheet correlators for ${\rm AdS}_3$,''
JHEP \textbf{02} (2021) 081
{\tt [\href{https://arxiv.org/abs/2009.11306}{arXiv:2009.11306 [hep-th]}]}.




\bibitem{Quella:2007hr}
T.~Quella and V.~Schomerus,
``Free fermion resolution of supergroup WZNW models,''
JHEP \textbf{09} (2007) 085
{\tt[\href{https://arxiv.org/abs/0706.0744}{arXiv:0706.0744 [hep-th]}]}







\bibitem{Gaberdiel:2018rqv}
M.R.~Gaberdiel and R.~Gopakumar,
``Tensionless string spectra on AdS$_{3}$,''
JHEP \textbf{05} (2018) 085
{\tt[\href{https://arxiv.org/abs/1803.04423}{arXiv:1803.04423 [hep-th]}]{}}.

\bibitem{Berkovits:1999im}
N.~Berkovits, C.~Vafa and E.~Witten,
``Conformal field theory of AdS background with Ramond-Ramond flux,''
JHEP \textbf{03} (1999) 018
{\tt{\href{https://arxiv.org/abs/hep-th/9902098}{arXiv:hep-th/9902098 [hep-th]}}}.


\bibitem{Rozansky:1992rx}
L.~Rozansky and H.~Saleur,
``Quantum field theory for the multivariable Ale\-xander-Conway polynomial,''
Nucl.\ Phys.\ B \textbf{376} (1992) 461-509 \newline
{\tt[\href{https://arxiv.org/abs/hep-th/9203069}{arXiv:hep-th/9203069 [hep-th]}]{}}.





\bibitem{Halpern:1989ss}
M.B.~Halpern and E.~Kiritsis,
``General Virasoro Construction on Affine G,''
Mod.\ Phys.\ Lett.\ A \textbf{4} (1989) 1373.



\bibitem{Gaberdiel:2007jv}
M.R.~Gaberdiel and I.~Runkel,
``From boundary to bulk in logarithmic CFT,''
J.\ Phys.\ A \textbf{41} (2008) 075402
{\tt[\href{https://arxiv.org/abs/0707.0388}{arXiv:0707.0388 [hep-th]}]}.


\bibitem{Raclariu:2021zjz}
A.M.~Raclariu,
``Lectures on Celestial Holography,'' \newline
{\tt \href{https://arxiv.org/abs/2107.02075}{arXiv:2107.02075 [hep-th]}{}}.

\bibitem{Gaberdiel:1998ps}
M.R.~Gaberdiel and H.~G.~Kausch,
``A Local logarithmic conformal field theory,''
Nucl.\ Phys.\ B \textbf{538} (1999) 631-658
{\tt[\href{https://arxiv.org/abs/hep-th/9807091}{arXiv:hep-th/9807091 [hep-th]}]{}}.





\end{thebibliography}
\end{document}